\documentclass[usedcolumn,usenatbib,usegraphicx]{mn2e}

\title[Accretion disc formation in Be/X-ray binaries]
  {Accretion disc formation around the neutron star in Be/X-ray binaries}
\author[Kimitake~Hayasaki and Atsuo~T.~Okazaki]
  {Kimitake~Hayasaki$^1$\thanks{E-mail: kimi@astro1.sci.hokudai.ac.jp}
   and Atsuo~T.~Okazaki$^2$ \\
  $^1$Devision of Physics, graduate school of science, Hokkaido
  University, Kitaku, Sapporo 060-0810, Japan.\\
  $^2$Faculty of Engineering, Hokkai-Gakuen University, Toyohira-ku, 
      Sapporo 062-8605, Japan.}
\date{Received ...,; accepted ...}

\pagerange{\pageref{firstpage}--\pageref{lastpage}} \pubyear{2004}

\def\LaTeX{L\kern-.36em\raise.3ex\hbox{a}\kern-.15em
    T\kern-.1667em\lower.7ex\hbox{E}\kern-.125emX}

\begin{document}

\label{firstpage}

\maketitle

\begin{abstract}

We study the accretion on to the neutron star in Be/X-ray binaries,
using a 3D SPH code and the data imported from a simulation by \citet{oka2} 
for a coplanar system with a short period ($P_{\rm orb}=24.3\ \rm{d}$) 
and a moderate eccentricity $(e=0.34)$, which 
targeted the Be/X-ray binary 4U\,0115+63. For simplicity, we adopt the
polytropic equation of state. We find that a time-dependent accretion
disc is formed around the neutron star regardless of the simulation
parameters. In the long term, the disc evolves via a two-stage
process, which consists of the initial developing stage and the later
developed stage. The developed disc is nearly Keplerian. In the short
term, the disc structure modulates with the orbital phase. The disc
shrinks at the periastron passage of the Be star and restores its
radius afterwards. The accretion rate on to the neutron star is also
phase dependent, but its peak is broader and much lower than that of
the mass-transfer rate from the Be disc, unless the polytropic
exponent is as large as 5/3. Our simulations show that the truncated
Be disk model for Be/X-ray binaries is consistent with the observed
X-ray behaviour of 4U\,0115+63.

\end{abstract}


\section{Introduction}
\label{sec:intro}

The Be/X-ray binaries represent the largest subclass of high-mass
X-ray binaries. About two-thirds of the identified systems fall into
this category. These systems consist of, generally, a neutron star and
a Be star with a cool ($\sim 10^{4}K$) equatorial disc, which is
geometrically thin and nearly Keplerian. The orbit is wide $(16\,{\rm
 d} \la P_{\rm{orb}} \la 243\,\rmn{d})$ and usually eccentric ($e\ga
0.3$) (\rm{e.g.,} \citealt{zi}).

Most of the Be/X-ray binaries show only transient activity in the
X-ray emission and are termed Be/X-ray transients. Be/X-ray transients
show periodical (Type I) outbursts, which are separated by the orbital
period and have the lumiocity of
$L_{\rmn{X}}=10^{36-37}\rmn{erg\,s}^{-1}$, and giant (Type II)
outbursts of $L_{\rmn{X}} \ga 10^{38} \rmn{erg\,s}^{-1}$ with no
orbital modulation. The transient nature in the X-ray activity of
Be/X-ray binaries is considered to result from the interaction between
the accreted material from the circumsteller disc of the Be star and
the rotating magnetized neutron star \citep*{st}. If the
accreted material is dense enough to make the magnetospheric radius
smaller than the corotation radius, the accretion on to the neutron
star causes a bright X-ray emission (the direct accretion
regime). Otherwise, the magnetospheric radius is larger than the
corotation radius, and the accretion is prevented by the centrifugal
inhibition. The system is then in quiescence (the propeller regime).

During the outbursts, the neutron stars in most Be/X-ray binaries show
 spin-up (e.g., 4U\,0115+63, GS\,1843-02, EXO\,2030+375 and XTE\,J1946+26) 
(\citealt{bi}; \citealt{fi2}; \citealt{wi1}; \citealt{wi2}), 
and some even show quasi-periodic oscillations
(QPOs) (e.g., A\,0535+26 and EXO\,2030+375) (\citealt{fi1}; \citealt{wi1}). 
These observational features strongly suggest the presence of
an accretion disc during the outbursts.

The accretion disc is also indicated to exist between the outbursts
in several systems (e.g., A\,0535+26) 
(\citealt{fi1}; see also \citealt{zi}).
On the other hand, \citet{cl} searched the optical/infrared
contribution from the accretion disc in A\,0535+26 during the X-ray
quiescent phase, and found no signature of the accretion disc in
propeller regime. Obviously, more work is needed to determine whether
the accretion disc also exists in the quiescent state.

Recently, \citet{oka2} studied the interaction between the
Be-star disc and the neutron star, using a three-dimensional (3D)
smoothed particle hydrodynamics (SPH) code (\citealt{be1}; \citealt*{ba}). 
They also qualitatively discussed the relation between the
X-ray luminosity and the accretion rate using the simulated
mass-capture rate by the neutron star. In their simulations, however,
the neutron star was modelled by a sink particle \citep{ba}
with the size of the Roche lobe. Hence, no information was available
how the captured material accretes on to the neutron star. Moreover,
no direct comparison with the observed X-ray data was possible.

In this paper, we study the accretion on to the neutron star in
Be/X-ray binaries, using a 3D SPH code. The simulation should ideally
take account of all the processes at work in a Be/X-ray binary, such
as the disc evolution around the Be star, the mass transfer from the
Be-star disc to the neutron star, and the accretion on to the neutron
star. Such a simulation, however, would require very intense
computations.

Therefore, we confine ourselves to simulate only the accretion flow
around the neutron star, with an outer boundary condition constructed
by the data of the particles captured by the neutron star from a
high-resolution simulation by \citet{oka2}. This situation is
illustrated in Fig.~1, where the right panel taken from their
simulation shows the mass transfer from the Be-star disc to the
neutron star at periastron, while the left panel shows the accretion
disc formed around the neutron star in one of our simulations.

\begin{figure}
\includegraphics*[width=8.5cm]{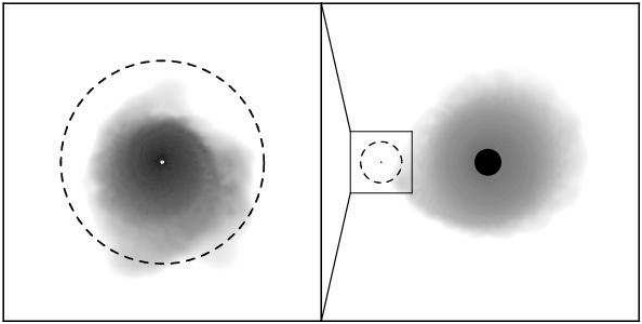} 
 \caption{Illustrative diagram of a Be/X-ray binary. At the periastron
   passage, the gas in an outer part of the Be-star disc is
   transferred to the neutron star and forms an  accretion disc. The
   right panel is taken from a simulation by \citet{oka2},
   while the left panel is from one of our simulations.}
 \label{figure1}
\end{figure}

\section{Numerical model}
\label{sec:model}

\subsection{SPH code}
\label{sec:sph}

Simulations presented here were performed with a 3D SPH code. The SPH
code is basically the same as that used by \citet{oka2},
which is based on a version originally developed by Benz (\citealt{be1}; \citealt{be2}). 
The smoothing length is variable in time and
space. The SPH equation with the standard cubic-spline kernel are
integrated using a second-order Runge-Kutta-Fehlberg integrator with
individual time steps for each particle \citep{ba},
which results in an enormous computational saving when a large range
of dynamical time-scales are involved.

In our code, the accretion flow on to the neutron star is modelled by
an ensemble of gas particles, each of which has an arbitrary and 
negligible mass chosen to be $10^{-15}M_{\sun}$ with a variable smoothing length.
Note that the accretion flow is nonself-gravitating in our simulations.
On the other hand, the Be star and the neutron star are modelled by two sink
particles \citep{ba} with corresponding masses. 
The gas particles which fall within a specified accretion radius are accreted
by the sink particles. We assume that the neutron star has the fixed
accretion radius of $3.0-5.0\times10^{-3}a$, $a$ being the semi-major
axis of the binary. For the Be star, we adopt a variable accretion
radius of $0.8r_{\rmn{L}}$, where $r_{\rmn{L}}$ is the Roche-lobe
radius for a circular binary. This is because our interest is only in
the flow around the neutron star. An approximate formula for
$r_{\rmn{L}}$ is given by
\begin{equation}
  r_{L}\simeq 0.462\left(\frac{1}{1+q}\right)^{1/3}D,
  \label{eq:roche}
\end{equation}
(see, for example, \citealt{wa}; \citealt*{fr}) with the mass ratio
$q=M_{\rmn{X}}/M_{*}$, where $M_{\rmn{X}}$ and $M_{*}$ are the masses
of the neutron star and the Be star, respectively, and $D$ is the
distance between the stars.

In disc simulations using a three dimensional SPH code with the
standard cubic spline kernel, one has an approximate relation between
the Shakura-Sunyaev viscosity parameter $\alpha_{\rmn{SS}}$ and the
linear SPH artificial viscosity parameter $\alpha_{\rmn{SPH}}$ as
\begin{equation}
  \alpha_{\rmn{SS}} = \frac{1}{10}\alpha_{\rmn{SPH}}\frac{h}{H},
  \label{eq:alopha_ss}
\end{equation}
if the non-linear SPH artifical viscosity parameter
$\beta_{\rm SPH}$ is set equal to zero \citep{oka2}. 
Here $h$ is the SPH smoothing length and 
$H$ is the scale-height of the disc.

In a simulation we report in this paper, 
we adopt $\alpha_{\rmn{SPH}}=1$ and $\beta_{\rmn{SPH}}=2$ 
in which case $\alpha_{\rmn{SS}}$ is variable in time and space. 
In the other simulations, however, we adopt a constant value of
$\alpha_{\rmn{SS}}$ in order to roughly model the $\alpha$ disc. In
these simulations, $\alpha_{\rmn{SPH}} =10 \alpha_{\rmn{SS}} H/h$ was
variable in time and space and $\beta_{\rmn{SPH}}=0$.

In our simulations, we adopt a polytropic equation of state specified
by the exponent $\Gamma$, in which the effects of an energy
dissipation resulting from viscosity and a radiative cooling are 
approximately boxed.

\subsection{Boundary condition}
\label{sec:bc}

Our binary configuration is the same as that of \citet{oka2}. 
We set the binary orbit on the $x$-$y$ plane with the
semi-major axis along the $x$-axis. Initially, the neutron star is at
the periastron. It orbits about the Be star with the orbital period
$P_{\rmn{orb}}=24.3\,\rmn{d}$ and the eccentricity $e=0.34$. 
The unit of time adopted here and throughout the paper is $P_{\rmn{orb}}$. 
The Be disc is coplanar with the binary orbital plane. As the Be star, we
take a B0V star with $M_{*} = 18 M_{\sun}$, $R_{*}=8R_{\sun}$ and
$T_{\rmn{eff}}=26000\,{\rmn{K}}$, where $R_{*}$ and $T_{\rmn{eff}}$
are the stellar radius and effective temprature, respectively. For the
neutron star, we take $M_{\rmn{X}} = 1.4 M_{\sun}$ and $R_{\rmn{X}} =
10^{6}\,\rmn{cm}$. These parameters are adopted to target 4U\,0115+63, 
which has shown only Type~II X-ray outbursts followed by a short series of
Type~I X-ray outbursts \citep{ne1}.

\begin{figure}
\centerline{\includegraphics[height=6cm,width=8cm,clip]{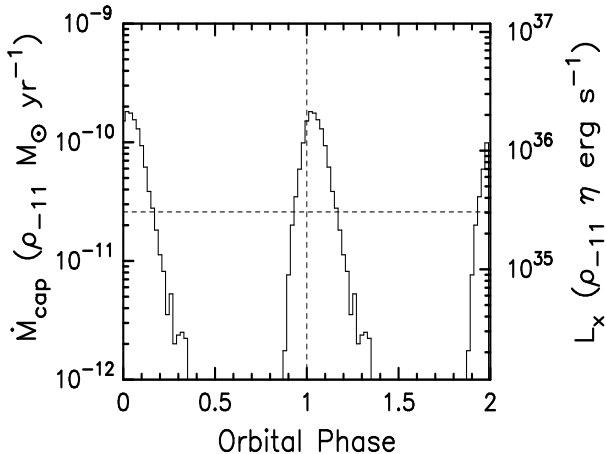}}
 \caption{
 Orbital-phase dependence of the mass-capture rate by the neutron
 star, taken from a high-resolution simulation by \citet{oka2}. 
 The data is folded on the orbital period over $32 \le t \le
 47$. The mass-capture rate is measured in units of $\rho_{-11}
 M_{\sun}\,{\rmn{yr}}^{-1}$, where $\rmn{\rho_{-11}}$ is highest local
 density in the Be-star disc normalized by $\rmn{10^{-11}} \rmn{g}
 \rmn{cm^{-3}}$, a typical value for Be stars. The right axis shows
 the X-ray luminosity corresponding to the mass-capture rate with the
 X-ray emission efficiency $\eta=1$, where $\eta$ is defined by
 $L_{X}=\eta M_{\rmn{X}}\dot{M}_{\rmn{cap}}/R_{\rmn{X}}$.}
 \label{fig:mdot_cap}
\end{figure}

As mentioned earlier, we concentrate only on the simulations of the
accretion flow on to the neutron star. This is done by constructing an
outer boundary condition from the data of particles captured by the
neutron star in a high-resolution simulation by \citet{oka2}, 
which was run for $0 \le t \le 47$ and finally had $\sim
140000$ particles with $\alpha_{\rmn{SS}}=0.1$ . We first fold the
data on the orbital period over $32 \le t \le 47$ to reduce the
fluctuation noise. The phase dependence of the mass-capture rate by
the neutron star obtained by this procedure is shown in
Fig.~\ref{fig:mdot_cap}. Note that there is a strong phase-dependence
in the mass-transfer rate. The peak comes slightly after the
periastron passage and decreases by two orders of magnitude by 
apastron passage. We then fit the spatial and velocity distribution of the
captured particles in each phase bin with separate Gaussian
functions. This is for running simulations with an arbitrary
mass-transfer rate from the Be disc. We finally model the
mass-transfer process from the Be disc to the neutron star by
injecting gas particles at a given phase-dependent rate shown in 
Fig. 2 with the spatial and velocity distributions specified by these
Gaussian functions. The injected particles are assumed to have the
initial temperature of $T_{\rmn{eff}}/2$.

As described earlier, the mass of each gas particle is 
arbitrary and negligible in our simulations. All the mass-related 
quantities discussed later, such as the disc mass 
and the mass-accretion rate, are normalised by using the mass-capture 
rate shown in Fig.~2.


\begin{figure}
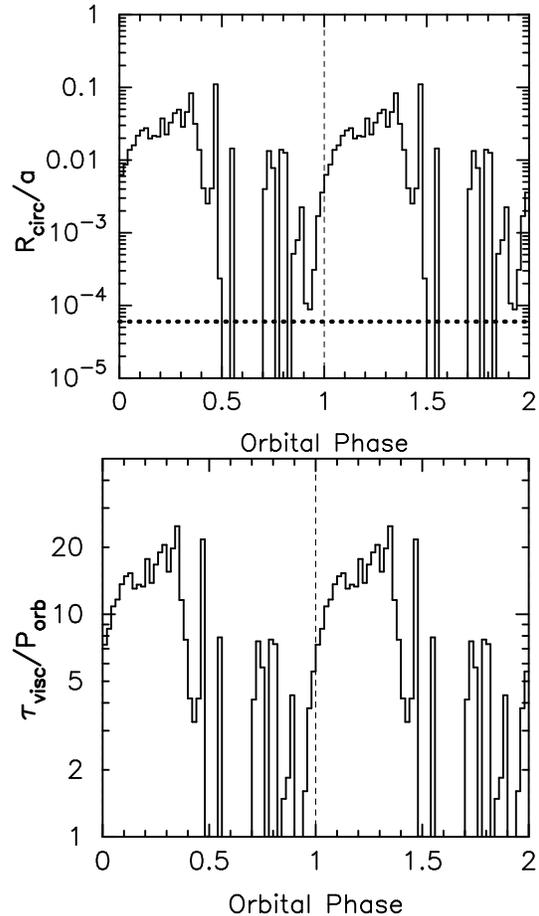

\centerline{
\includegraphics*[width=7.0cm,clip]{khfig3.ps}}
\centerline{
\includegraphics*[width=7.0cm,clip]{khfig4.ps}}
 \caption{Orbital-phase dependence of the circularization radius (the
   upper panel) and the ratio of the viscous time-scale to the orbital
   period (the lower panel) for the gas transferred from the Be disc
   in 4U\,0115+63, based on the data imported from a simulation by \citet{oka2}. 
   The horizontal dotted line in the upper
   panel denotes the corotation radius
   $R_{\rmn{\Omega}}=6.0\times10^{-5}a$. }
 \label{fig:rcirc}
\end{figure}


\begin{table*}
\caption{Summary of models of our simulations. The initial number of
  particles is 301 and the mean mass injection rate is
  $2.5\times10^{-11}\rho_{-11}M_{\odot}\rmn{yr}^{-1}$ in all
  simulations. The first column represents the model. The second
  column is the polytropic exponent $\Gamma$ and the third column is
  the number of SPH particles at the end of the run, which is given in
  units of $P_{\rmn{orb}}$ in the fourth column. The viscosity
  parameters adopted are given in the fifth column and the sixth
  column is the inner radius of the simulation region. The seventh
  column is the number of injected particles for one orbital
  period. The last two columns are the number of accreted particles
  for the last one orbital period and the corresponding mean accretion
  rate, respectively.
}
\label{tbl:models}
\begin{tabular}{@{}lcccccccc}
\hline
Model       & Polytropic exponent & $N_{\rmn{SPH}}$ 
            & Run time        & Viscosity parameters 
            & Inner boundary & $N_{\rmn{inj}}$
            & $N_{\rmn{acc}}$ 
            & $\dot{M}_{\rmn{acc}}$ \\
            & $\Gamma$        &    (final)
            & $(P_{\rmn{orb}})$   & 
            & $r_{\rmn{in}}/a$ & 
            & 
            & $(\rho_{-11}M_{\odot}\rmn{yr}^{-1})$ \\
\hline
1           & $1.2$ & $45687$
            & $8$  & $\alpha_{\rmn{SS}}=0.1$
            & $3.0\times10^{-3}$ & 8657 
            & 2990
            & $8.3\times10^{-12}$\\
2           & $1.2$ & $35821$
            & $8$   &  $\alpha_{\rmn{SS}}=0.1$ 
            & $5.0\times10^{-3}$ & 8717
            & 4054
            & $1.1\times10^{-11}$\\
3           & $1\hspace{1mm}(\rmn{isothermal})$ & $39295$
            & $8$   & $\alpha_{\rmn{SS}}=0.1$  
            & $5.0\times10^{-3}$  & 8844
            & 2676 
            & $7.4\times10^{-12}$\\
4           & $5/3\hspace{1mm}(\rmn{adiabatic})$ & $20118$
            & $8$   & $\alpha_{\rmn{SS}}=0.1$
            & $5.0\times10^{-3}$ & 8594
            & 6851
            & $1.9\times10^{-11}$\\
5           & $1.2$ & $47730$
            & $8$   & $\alpha_{\rmn{SPH}}=1, \beta_{\rmn{SPH}}=2$
            & $5.0\times10^{-3}$ & 8705
            & 2671
            & $7.5\times10^{-12}$ \\
6           & $1.2$ & $17488$
            & $1$   & $\alpha_{\rmn{SS}}=0.1$   
            & $5.0\times10^{-3}$ & 34778 
            & 16829
            & $1.2\times10^{-11}$ \\           
\hline
\end{tabular}
\end{table*}

\subsection{Possibility of an accretion disc formation}
\label{sec:possibility}

Before describing the results of our simulations, we discuss below the
possibility of the formation of an accretion disc around the neutron
star.

We first consider the radius $R_{\rmn{circ}}$ at which the material
transferred from the Be disc is circularized via the interaction with
the previously transferred gas. We assume the circularization radius
$R_{\rmn{circ}}$ to be approximately given by
$R_{\rm{circ}}=J^{2}/GM_{\rm{X}}$, where $J$ is the initial specific
angular momentum of the transferred material.

In the upper panel of Fig.~\ref{fig:rcirc}, we show the dependence of
$R_{\rmn{circ}}$ on the orbital phase, based on the data taken from a 
simulation by \citet{oka2}. It is noted that there is a
strong phase-dependence of $R_{\rmn{circ}}$. It has a minimum of
$\sim10^{-4}a$ slightly before periastron and then rapidly increases
with phase. For comparison, we also show the corotation radius,
$R_{\rmn{\Omega}}$, for 4U\,0115+63 by the dotted line. Here,
$R_{\rmn{\Omega}}$ is given by
\begin{equation}
  R_{\rmn{\Omega}}
  =\left(\frac{GM_{\rm{X}}P_{\rmn{spin}}^{2}}{4\pi^{2}}\right)^{1/3}
  =1.5 \times 10^{8} P_{\rmn{spin}}^{2/3} \left(
  \frac{M_{\rm{X}}}{M_{\odot}} \right)^{1/3}\,{\rmn{cm}},
  \label{eq:corotation}
\end{equation}
where $P_{\rmn{spin}}=3.61\ \rmn{s}$ is the spin period of the neutron
star in 4U\,0115+63. Note that $R_{\rmn{circ}} \gg R_{\rmn{\Omega}}$
for most of the orbital phase. Therefore, we expect that an accretion
disc is formed around the neutron star in 4U\,0115+63.

Next, we discuss whether the accretion disc formed around the neutron
star is persistent or transient. This is done by comparing the viscous
accretion time-scale with the orbital period. For simplicity, we assume
the accretion disc to be geometrically thin and isothermal with the
Shakura-Sunyaev viscosity parameter $\alpha_{\rmn{SS}}$ and evaluate
the accretion time-scale at $r=R_{\rmn{circ}}$. The ratio of these two
time-scales for 4U\,0115+63 is then given by
\begin{eqnarray}
  \frac{\tau_{\rm{vis}}}{P_{\rm{orb}}}
  &\sim&\frac{1}{2\pi\alpha_{\rm{SS}}c_{\rmn{s}}^{2}}
  \left( \frac{R_{\rmn{circ}}}{a} \right)^{1/2} \frac{GM_{\rm{X}}}{a}
  \left( 1+\frac{1}{q} \right)^{1/2}, \cr
  &\sim&1.2\times10^{2}\left( \frac{\alpha_{\rm{SS}}}{0.1}
  \right)^{-1}
  \left[ \frac{R_{\rmn{circ}}}{a} \right]^{1/2} \left(
  \frac{T_{\rm{d}}}{10^{4}\rm{K}} \right)^{-1},
  \label{eq:tratio}
\end{eqnarray}
where $c_{\rmn{s}}$ is the sound speed and $T_{\rmn{d}}$ is the disc
temprature. The orbital-phase dependence of
$\tau_{\rmn{vis}}/P_{\rmn{orb}}$ is shown in the lower panel of
Fig.~\ref{fig:rcirc}. It is immediately observed that the viscous
accretion time-scale is much longer than the orbital period for 
most of the orbital phase. Therefore, once formed, the accretion disc
around the neutron star in 4U\,0115+63 will survive over the whole
orbital phase. Note that this result depends little on the orbital
period, because $R_{\rm{circ}}$ increases almost linearly with increasing $a$.

\section{Accretion flow around the neutron star}
\label{sec:results}

Based on the procedure described in the previous section, we have
performed several simulations with different parameters. Details of
the models are given in Table~\ref{tbl:models}. The polytropic
exponents considered are $\Gamma=1$ (model~3), 1.2 (models~1, 2, 5 and
6) and 5/3 (model~4). We have taken the inner radius of the simulation
region, $r_{\rmn{in}}$, to be $5.0\times10^{-3} a$ in all simulations
but model~1 with a smaller inner radius of
$r_{\rmn{in}}=3.0\times10^{-3} a$, which was run to study the effect
of the inner radius on the simulation results. In models~1-4 and 6, we
have assumed $\alpha_{\rmn{SS}}=0.1$, where $\alpha_{\rmn{SPH}}$
varies in time and space and $\beta_{\rmn{SPH}}=0$. For comparison
purpose, we have also run model~5 with $\alpha_{\rmn{SPH}}=1$ and
$\beta_{\rmn{SPH}}=2$, where $\alpha_{\rmn{SS}}$ varies in time and
space. Model~6 was run solely for clearly illustrating the initial
build-up phase of an accretion disc. For this purpose, we have taken
the injection rate of the SPH particles in model~6 four times as high
as in other models and run only for one orbital period.

Analysing multi-wavelength long term monitoring observations of
4U\,0115+63, \citet{ne1} found that the Be star undergoes
quasi-cyclic $(\sim3-5\rmn{yr})$ activity, losing and reforming its
circumsteller disc. Then, the mass transfer to the neutron star should
vary, depending on the state of the Be disc. By the time the Be disc
is fully developed and becomes capable of supplying mass to the
neutron star, even a large accretion disc would be
exhausted. Therefore, to investigate the evolution of an accretion
flow around the neutron star after the Be disc is fully developed, 
we performed 3D SPH simulations based on the numerical models
described above, where there is intially no gas particle around the
neutron star. Since all models have shown similar structure and
evolution of the accretion flow, we will mainly show the results from
model~1 with $\Gamma=1.2$ and $r_{\rmn{in}}= 3 \times 10^{-3} a$.

\subsection{Initial phase of the disc formation}
\label{sec:init_phase}

\begin{figure*}
\resizebox{\hsize}{!}
{
\includegraphics*{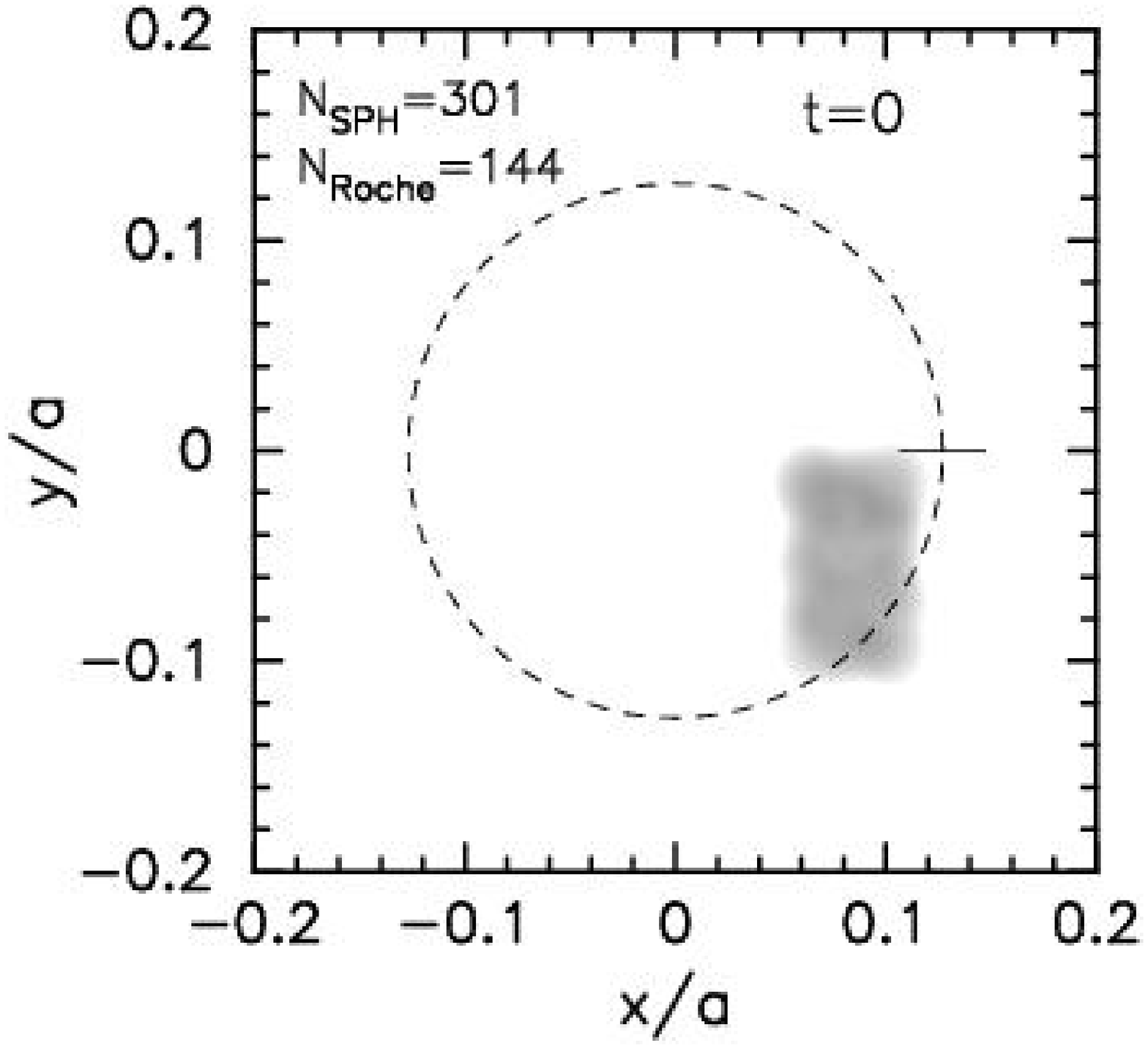}\hspace*{0.5em}
\includegraphics*{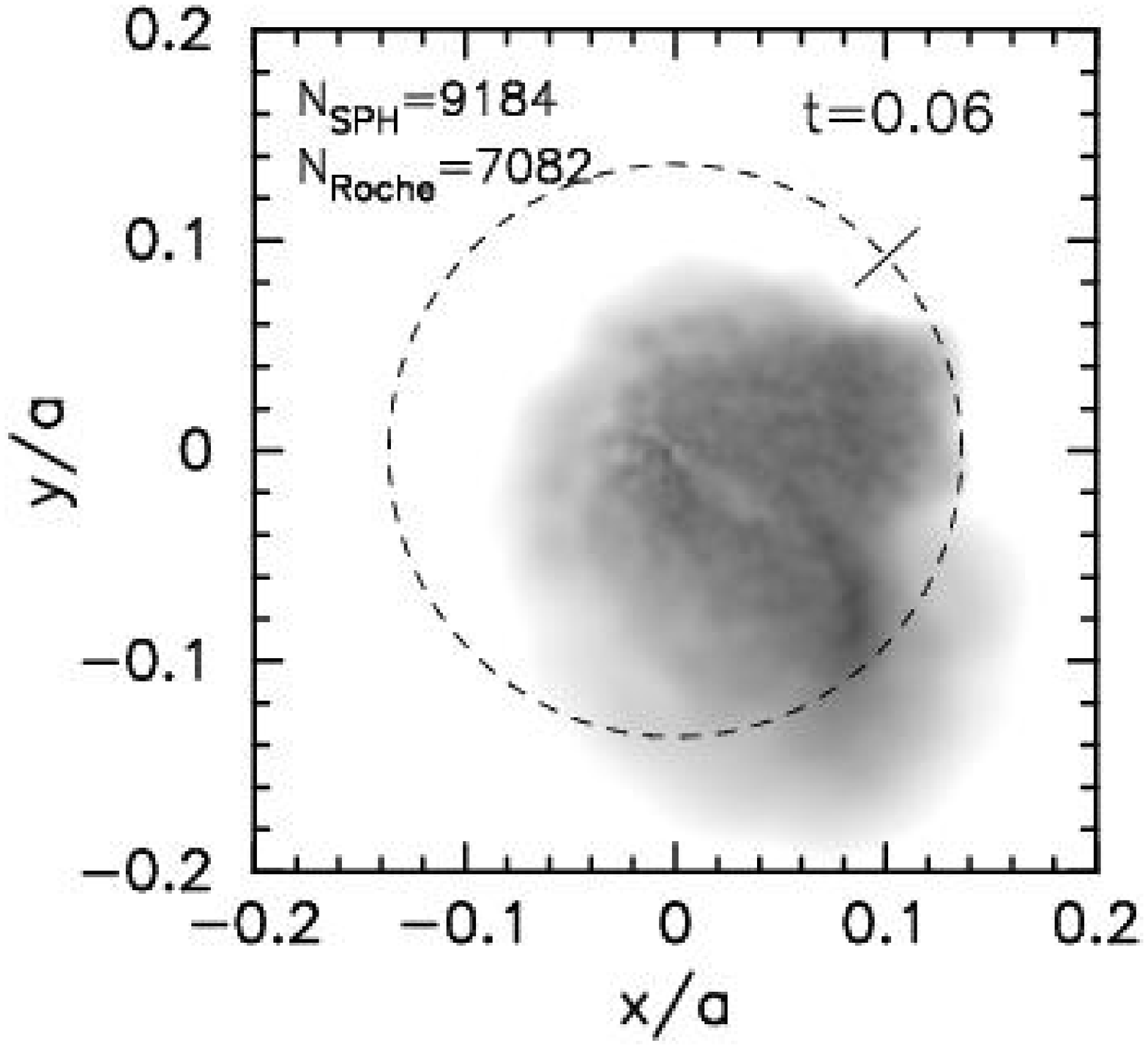}\hspace*{0.5em}
\includegraphics*{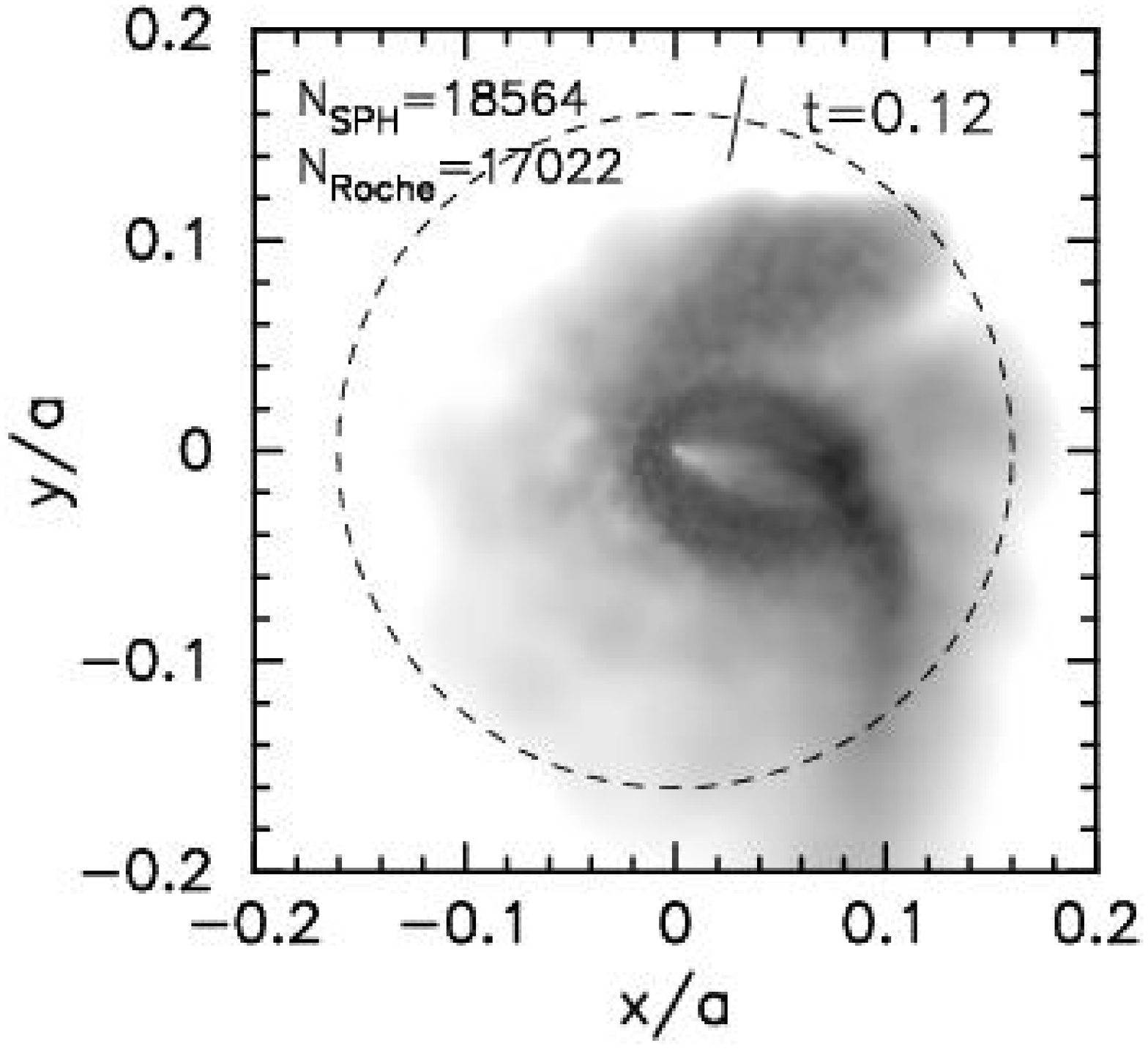}\hspace*{0.5em}
\includegraphics*{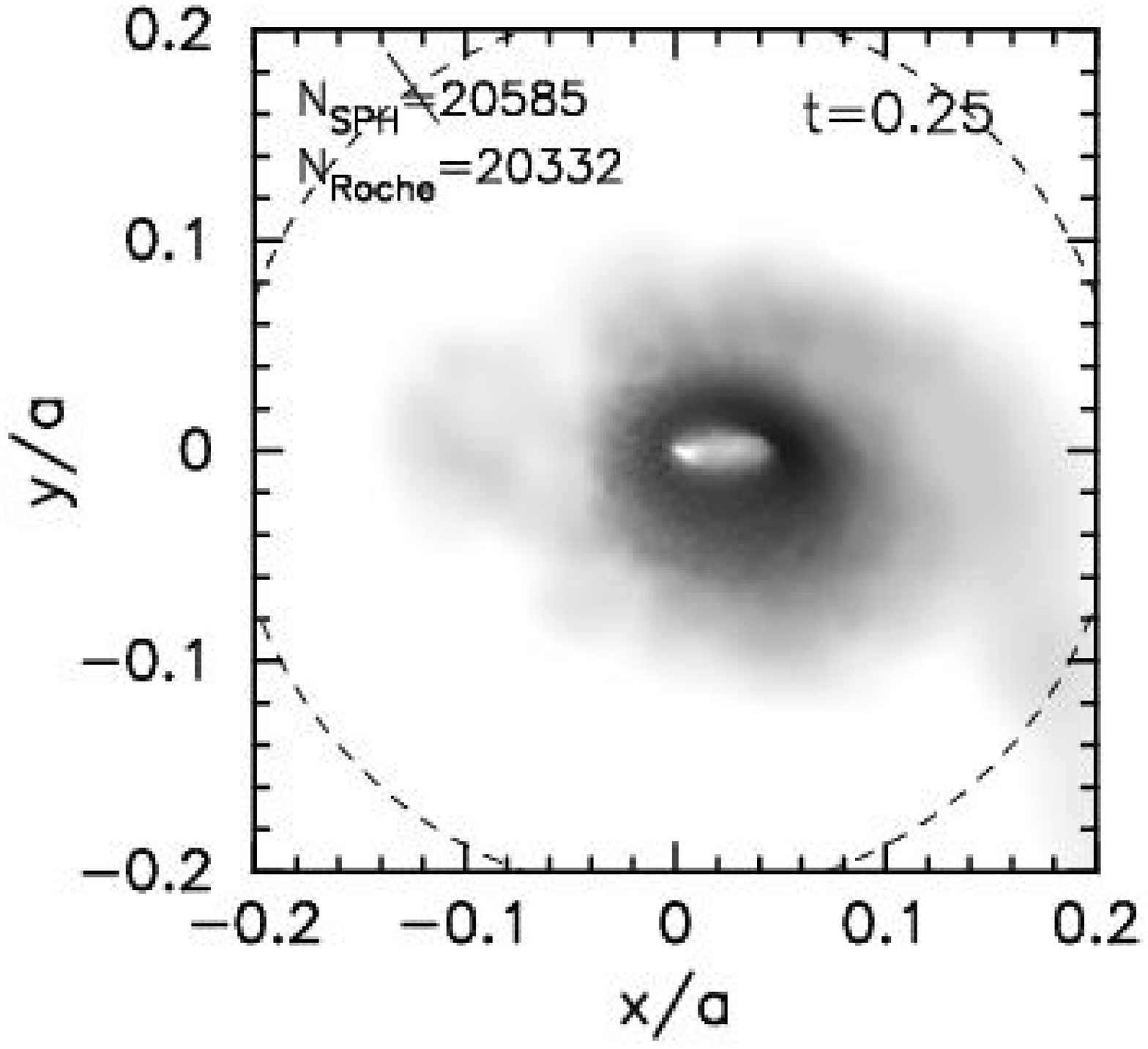}}\\
\resizebox{\hsize}{!}
{
\includegraphics*{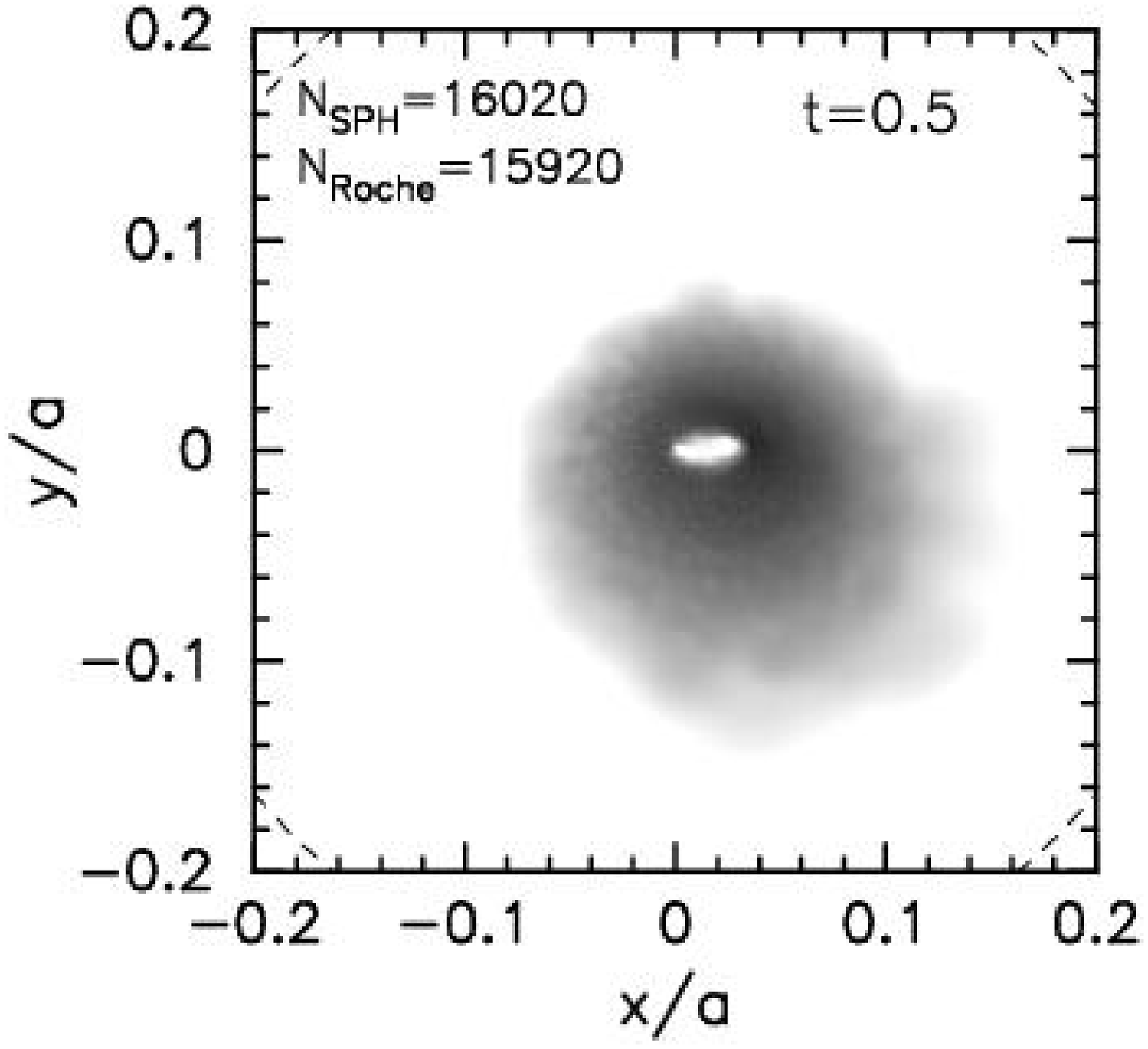}\hspace*{0.5em}
\includegraphics*{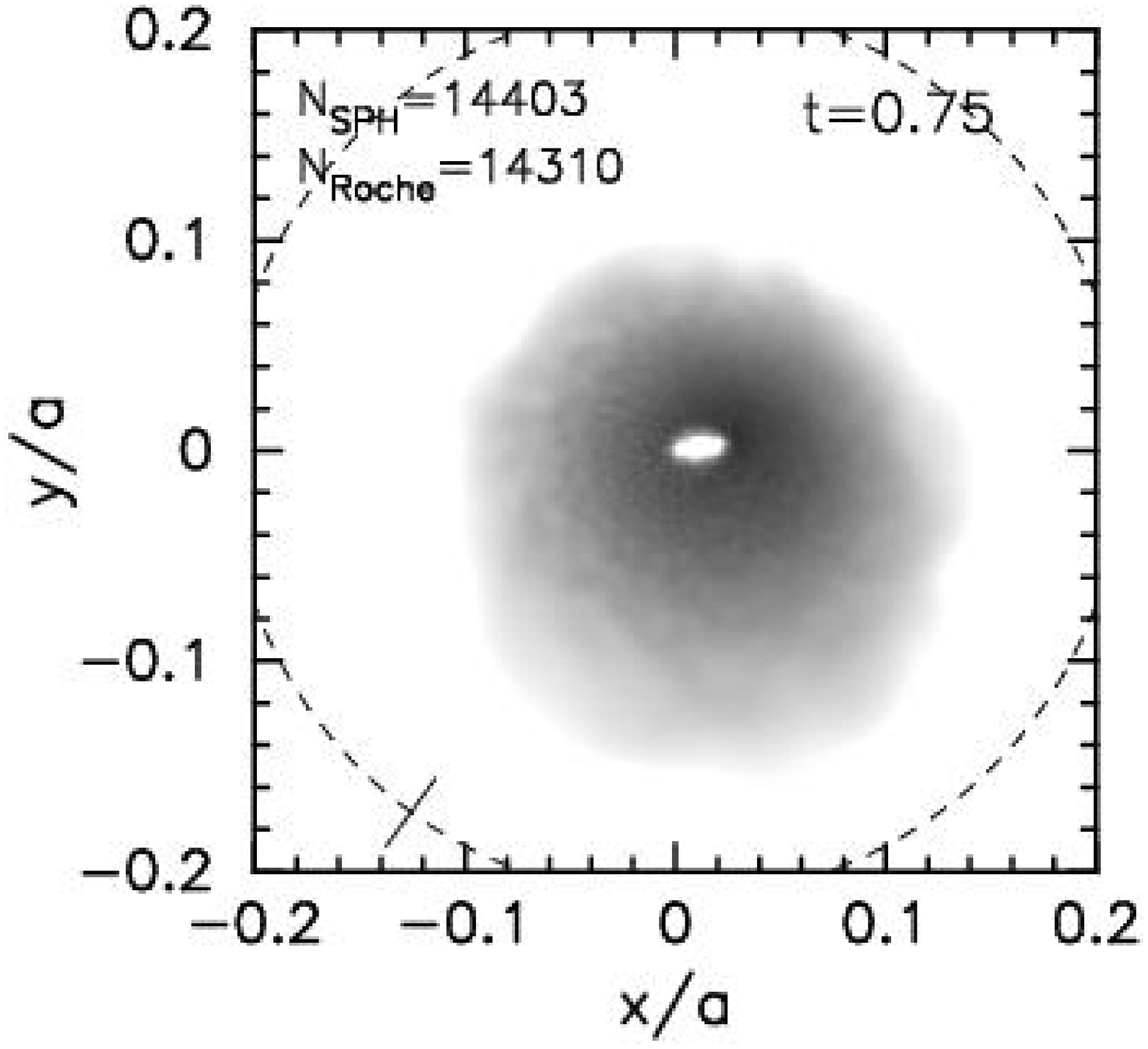}\hspace*{0.5em}
\includegraphics*{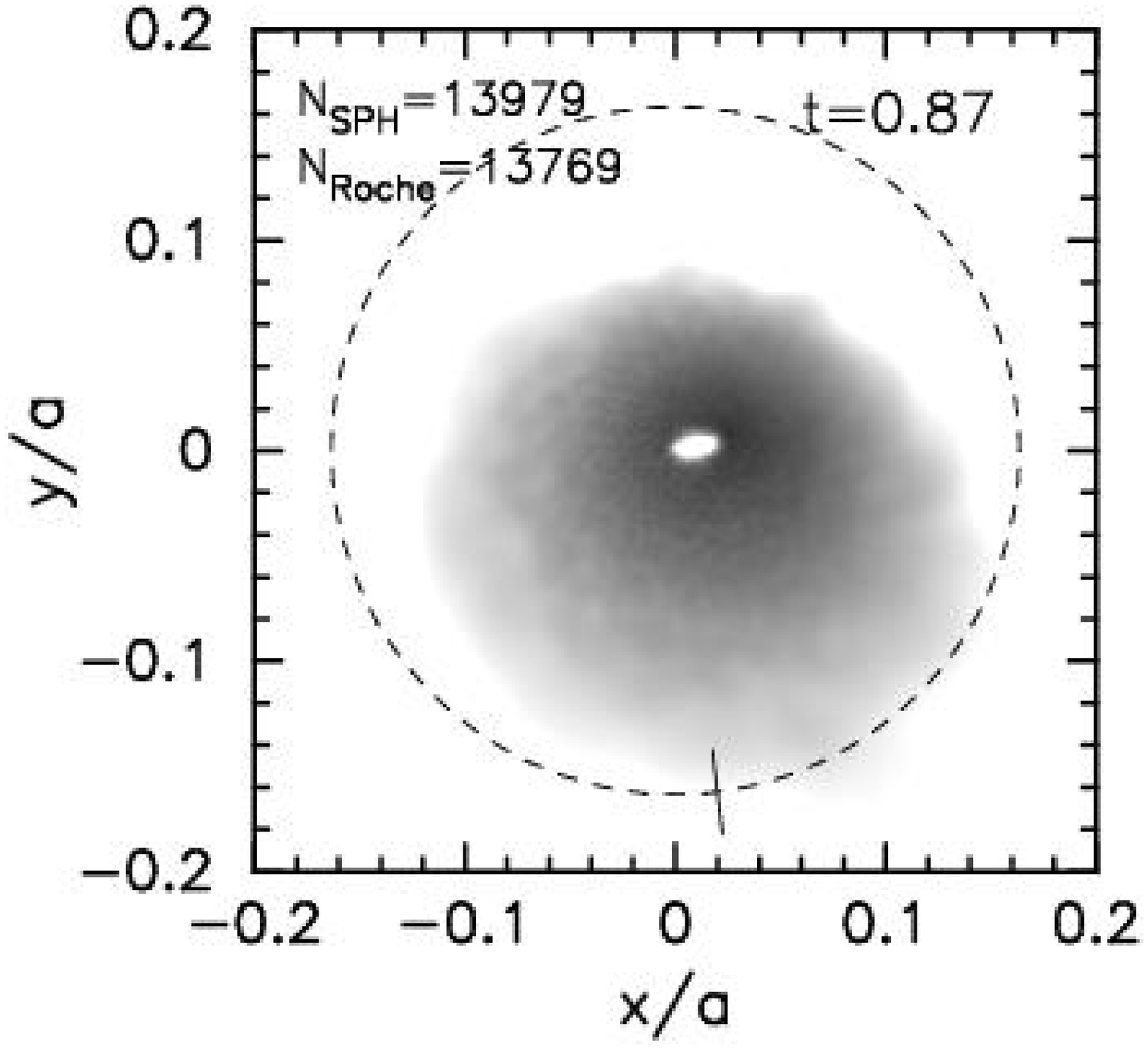}\hspace*{0.5em}
\includegraphics*{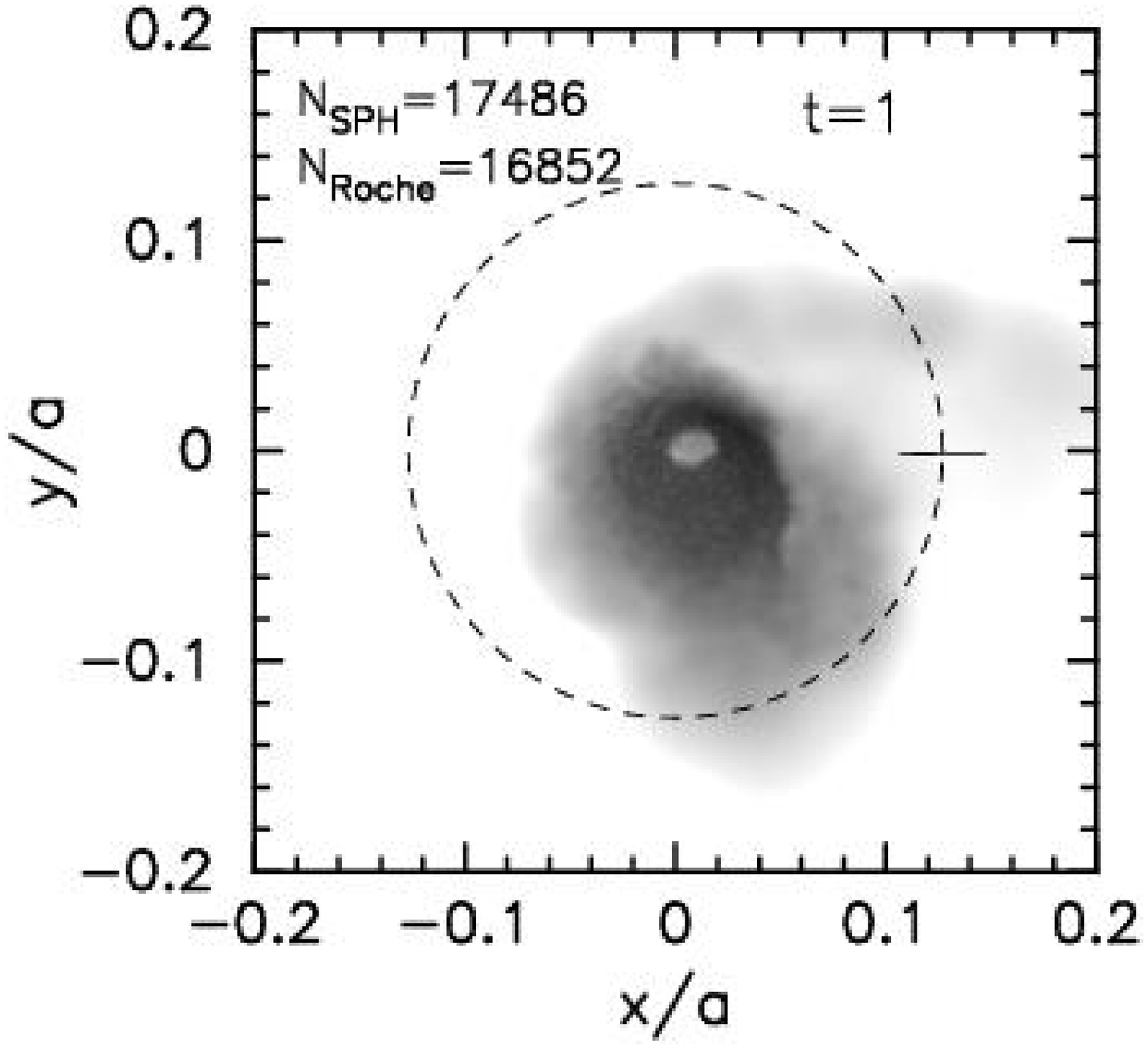}
}
 \caption{A sequence of snapshots showing the initial phase ($t=0-1$)
   of the disc formation around the neutron star in model~6 with
   $\Gamma=1.2$ and $r_{\rmn{in}}=5 \times 10^{-3} a$. The origin is
   set on the neutron star. Each panel shows the surface density in a
   range of four orders of magnitude in the logarithmic scale. The
   dashed circle denotes the effective Roche lobe of the neutron
   star. The short line-segment indicates the direction of the Be
   star. The unit of time is the orbital period
   $P_{\rmn{orb}}$. Annotated at the top-left corner of each panel are 
   the number of SPH particles, $N_{\rmn{SPH}}$, and the number of
   particles inside the effective Roche lobe of the neutron star,
   $N_{\rmn{Roche}}$.}
 \label{fig:model6_ss}
\end{figure*}

\begin{figure*}
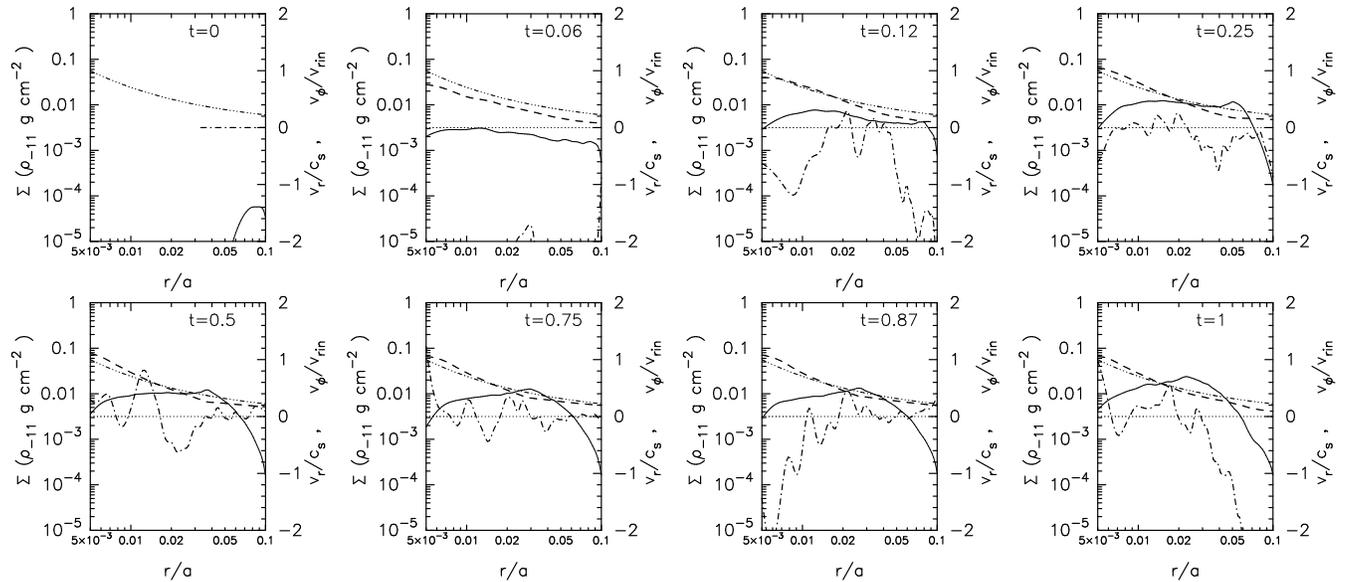

\resizebox{\hsize}{!}
{
\includegraphics*[width=4.5cm,clip]{khfig13.ps}\hspace*{0.5em}
\includegraphics*[width=4.5cm,clip]{khfig14.ps}\hspace*{0.5em}
\includegraphics*[width=4.5cm,clip]{khfig15.ps}\hspace*{0.5em}
\includegraphics*[width=4.5cm,clip]{khfig16.ps}}\\
\resizebox{\hsize}{!}
{
\includegraphics*[width=4.5cm,clip]{khfig17.ps}\hspace*{0.5em}
\includegraphics*[width=4.5cm,clip]{khfig18.ps}\hspace*{0.5em}
\includegraphics*[width=4.5cm,clip]{khfig19.ps}\hspace*{0.5em}
\includegraphics*[width=4.5cm,clip]{khfig20.ps}}
 \caption{Time-dependence of the radial structure of the
   accretion flow shown in Fig.~\ref{fig:model6_ss}. In each panel, the
  solid line, the dashed line and the dash-dotted line denote the
  surface density $\Sigma$ in units of
  $\rho_{-11}\,{\rmn{g\,cm}}^{-2}$, the azimuthal velocity $v_\phi$
  normalized by the Keplerian velocity at the inner boundary and the
  radial Mach number $v_r/c_{\rmn{s}}$, respectively. For comparison,
  the Keplerian velocity distribution is shown by the
  dash-three-dotted line. The density is integrated vertically and
  averaged azimuthally, while the velocity components are averaged
  vertically and azimuthally.}
 \label{fig:model6_rad}
\end{figure*}


We consider the initial phase of accretion flow around the neutron star.
As discussed in Section~\ref{sec:possibility}, it is expected that the
material transferred from the Be disc near the periastron does not
directly fall on to the neutron star and forms a disc-like structure
with a size corresponding to its specific angular momentum.

For illustration purpose, we give in Fig.~\ref{fig:model6_ss} the
snapshots of the initial accretion flow around the neutron star from
model~6, which covers $t=0-1$. At $t=0$, the Be star is at periastron. 
Each panel shows the surface density in the range of four orders of
magnitude in the logarithmic scale. We note that the accreting matter
first forms an elliptical ring around the neutron star, which is then
relaxed to an elliptical disc via the viscous diffusion. 
Because the viscous time-scale is longer than the orbital period, 
the orbits of
the accreting particles are not circularized and the flow is highly
eccentric during the initial phase of accretion. Comparing the number
of particles, $N_{\rmn{SPH}}$, with the number of particles inside the
effective Roche radius of the neutron star, $N_{\rmn{Roche}}$, we also
note that most of the material transferred from the Be disc stays
inside the effective Roche radius of the neutron star. It should be
noted that these dynamical features are common to all models studied
in this paper.

Fig.~\ref{fig:model6_rad} shows the time-dependence of the
radial structure of the accretion flow around the neutron star at the
same times as in Fig.~\ref{fig:model6_ss}. In each panel, the solid
line, the dashed line and the dash-dotted line denote the surface
density $\Sigma$, the azimuthal velocity $v_\phi$ normalized by the
Keplerian velocity at the inner boundary and the radial Mach number
$v_r/c_{\rmn{s}}$, respectively. In the figure, the density is
integrated vertically and averaged azimuthally, while the velocity
components are averaged vertically and azimuthally. From
Fig.~\ref{fig:model6_rad} we note that while the azimuthal velocity
distribution gradually approaches the Keplerian one, the behaviour of
the radial velocity is very complex, reflecting the dynamical process
of the disc formation shown in Fig.~\ref{fig:model6_ss}.

The surface density distribution also shows the initial evolution of
the accretion flow described above. In the panel at $t = 0.12$, there
appears a hump at $r \sim 0.08 a$ caused by the infall of injected
particles. Outside the hump the density decreases rapidly, making a
ring-like structure with a sharp outer edge. This rapid decrease in
the density is due to the ram pressure of the infalling particles. As
the hump moves inward, the density gradient outside the hump becomes
less steep. By the next periastron passage ($t = 1$), a disc-like
structure is formed, the inner radius of which is at $r \simeq 0.02
a$.

In our simulations, the presence of the inner simulation boundary
causes an artificial decrease in the density near the boundary,
because the gas particles which pass through the boundary are removed
from the simulation. As a result, the density distribution has a break
near the inner boundary, as seen in Fig.~\ref{fig:model6_rad}. 
We have found that our simulations are reliable for $\ga 2 r_{\rmn{in}}$.


\begin{figure*}
\resizebox{\hsize}{!}
{
\includegraphics*[width=4.5cm,clip]{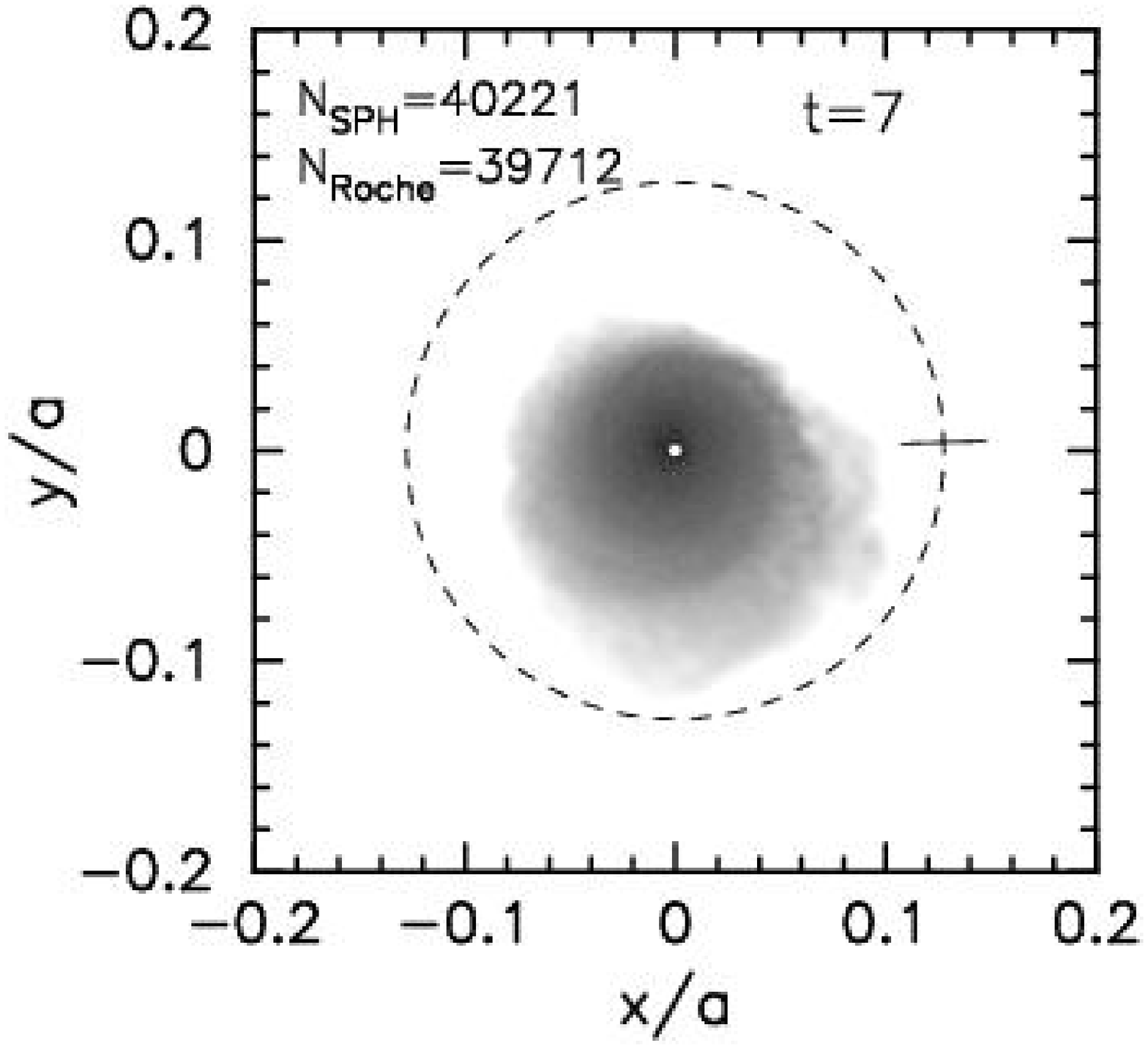}\hspace*{0.5em}
\includegraphics*[width=4.5cm,clip]{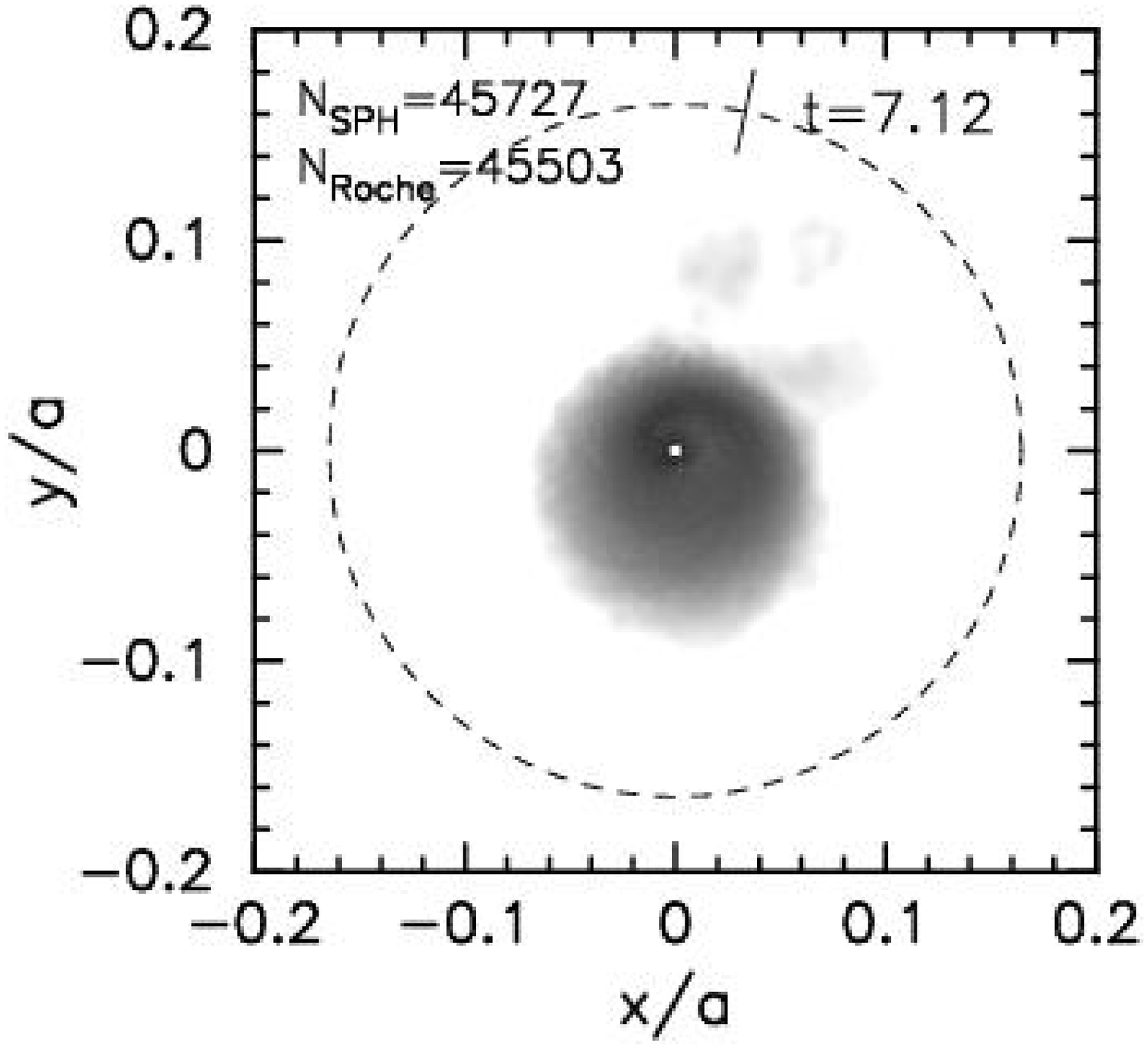}\hspace*{0.5em}
\includegraphics*[width=4.5cm,clip]{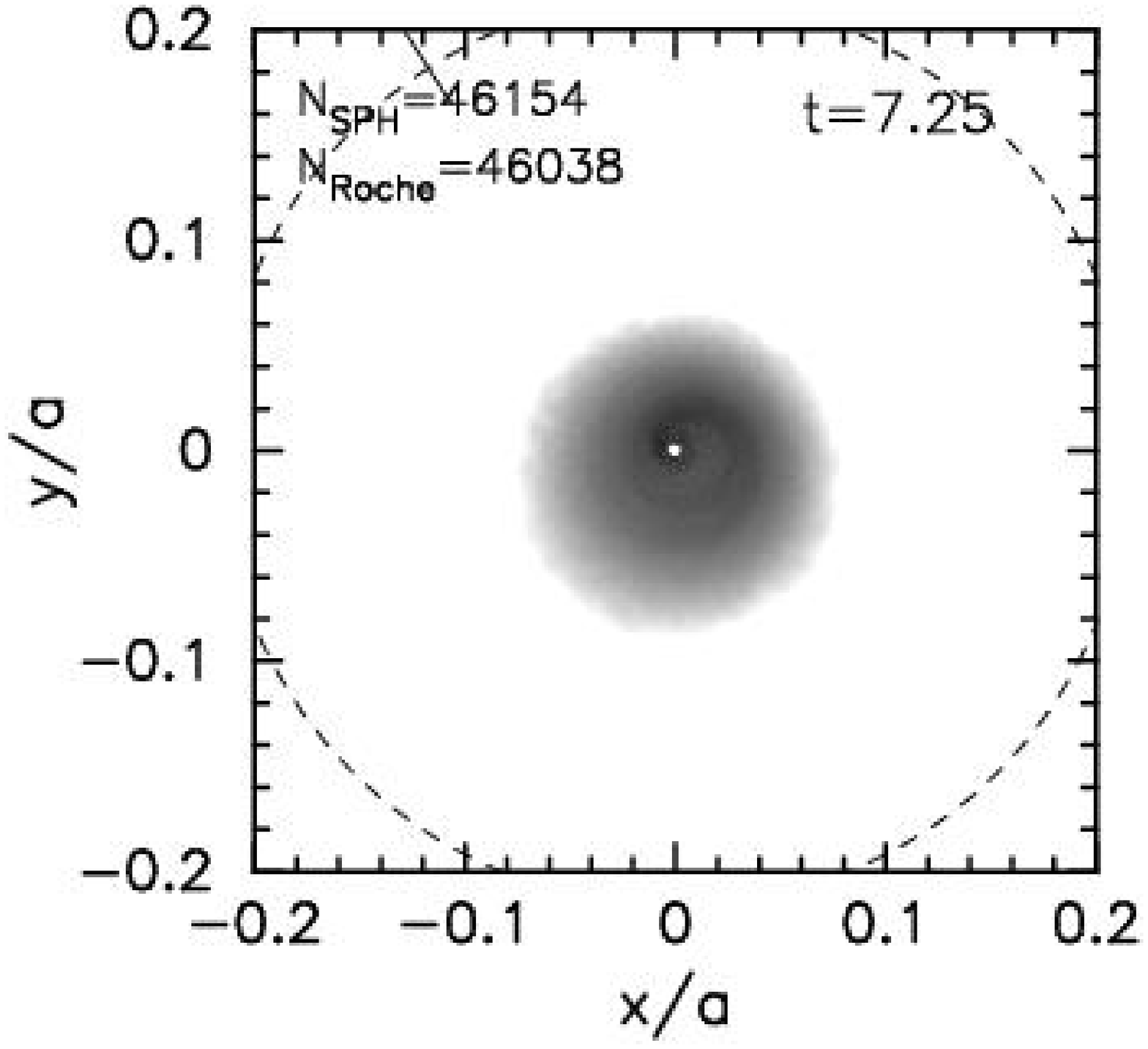}\hspace*{0.5em}
\includegraphics*[width=4.5cm,clip]{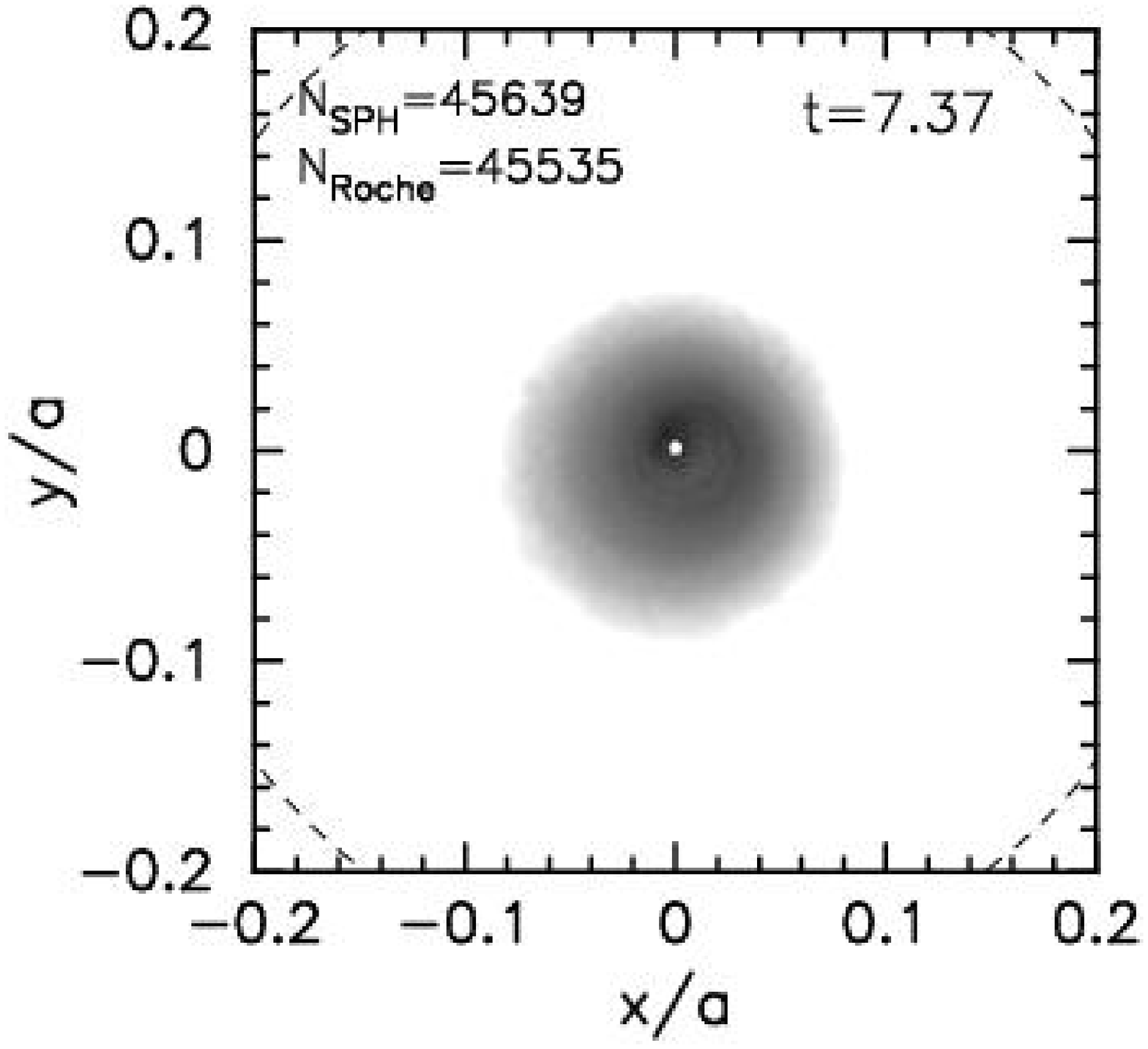}}\\
\resizebox{\hsize}{!}
{
\includegraphics*[width=4.5cm,clip]{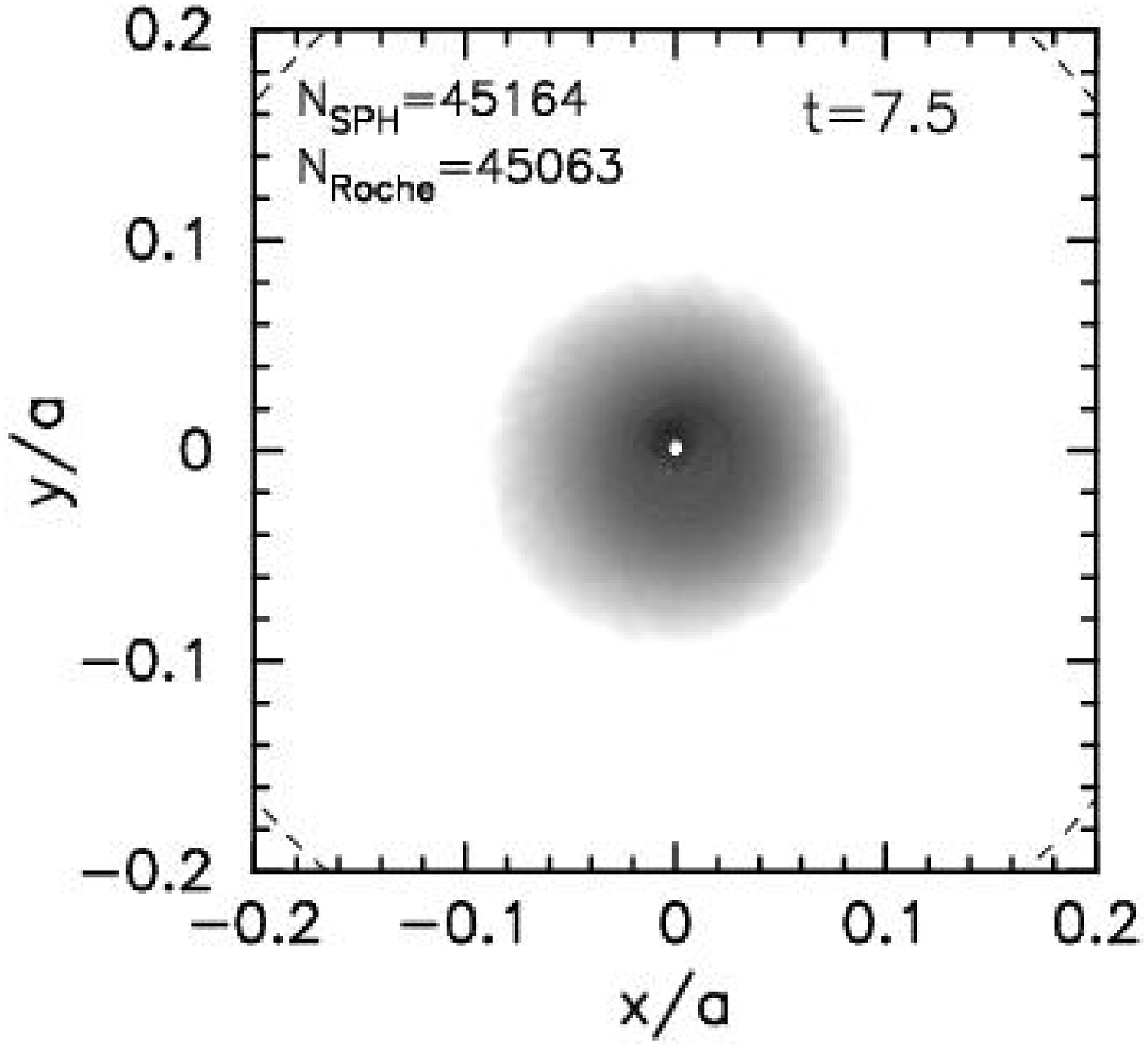}\hspace*{0.5em}
\includegraphics*[width=4.5cm,clip]{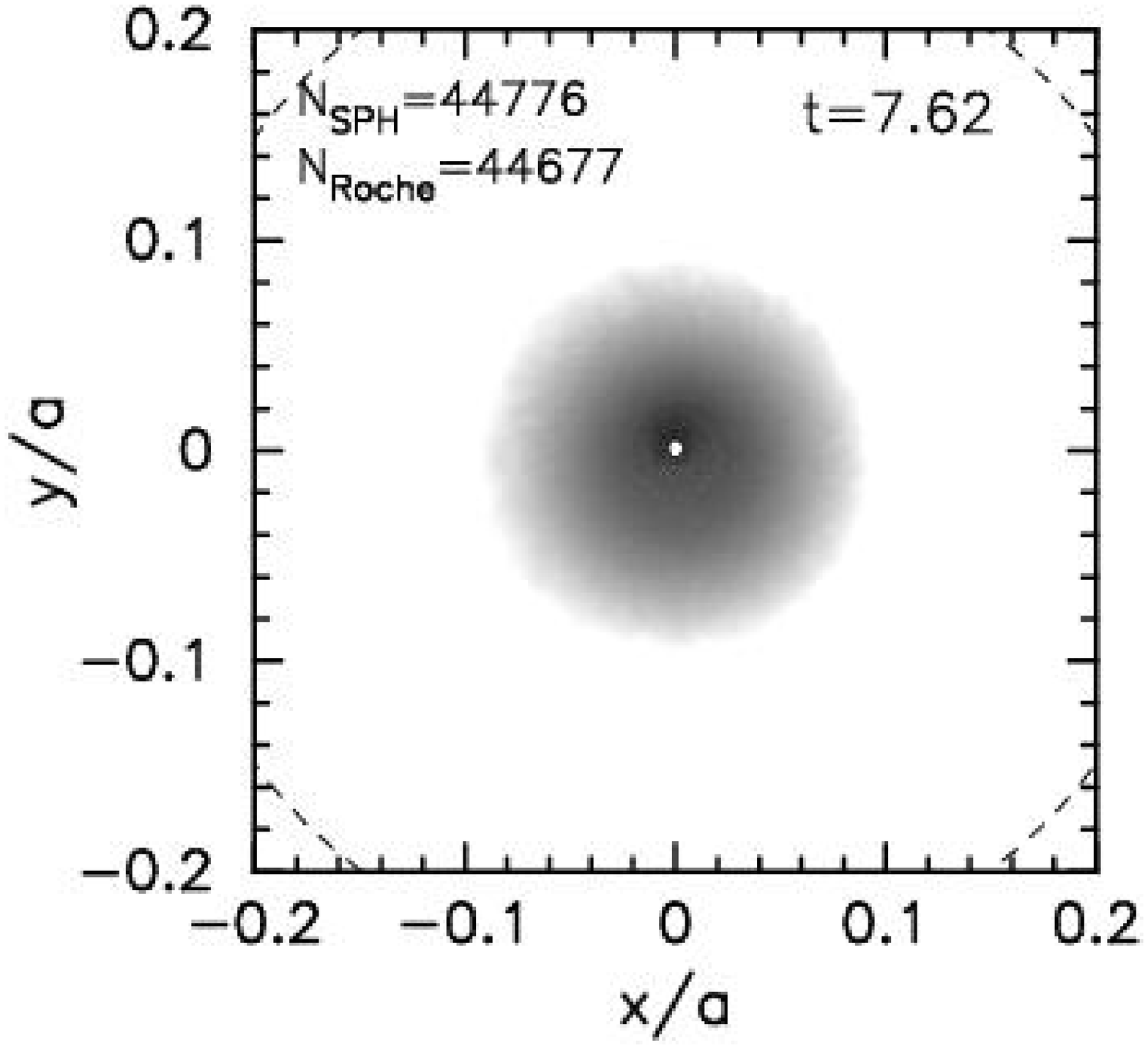}\hspace*{0.5em}
\includegraphics*[width=4.5cm,clip]{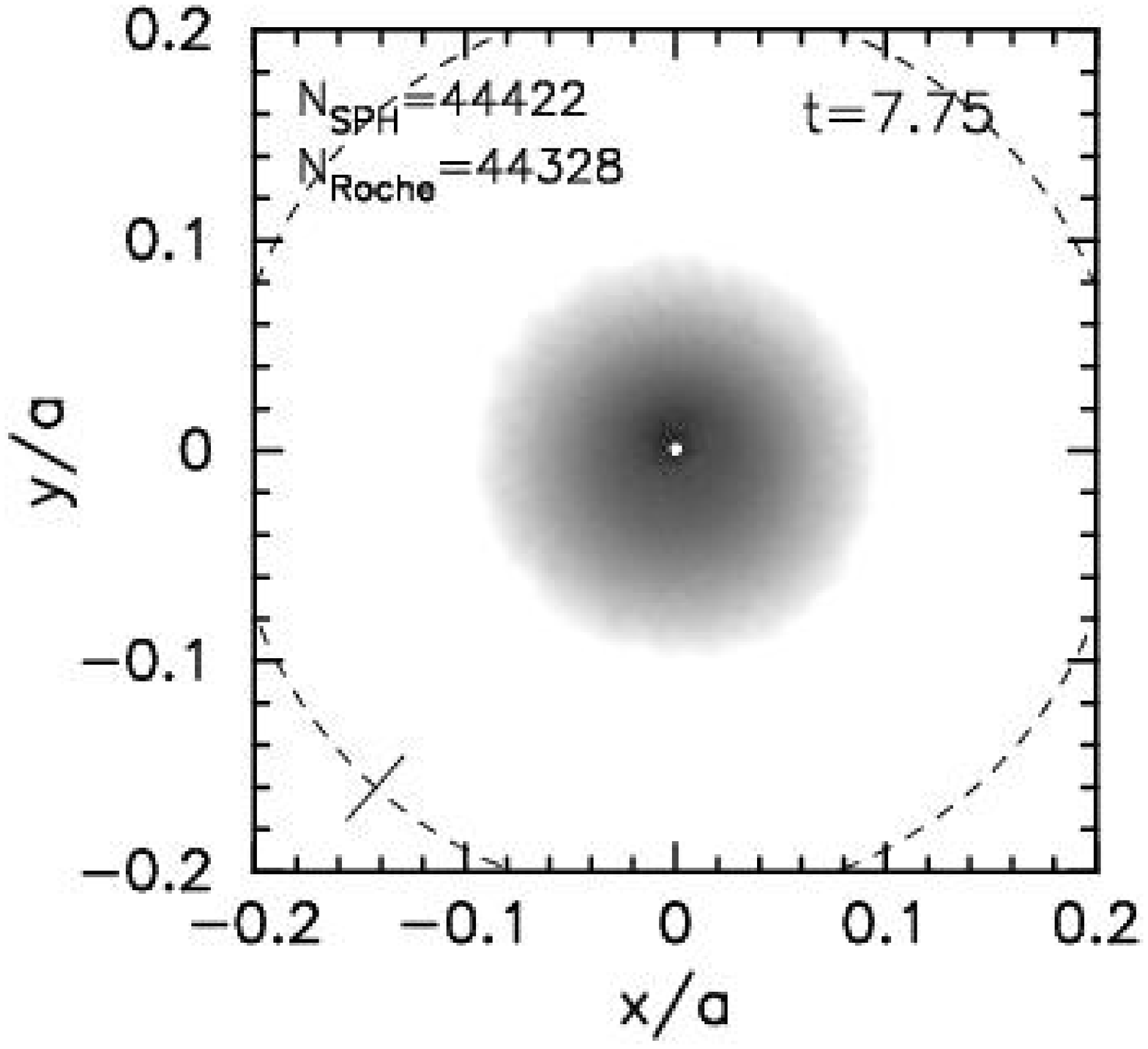}\hspace*{0.5em}
\includegraphics*[width=4.5cm,clip]{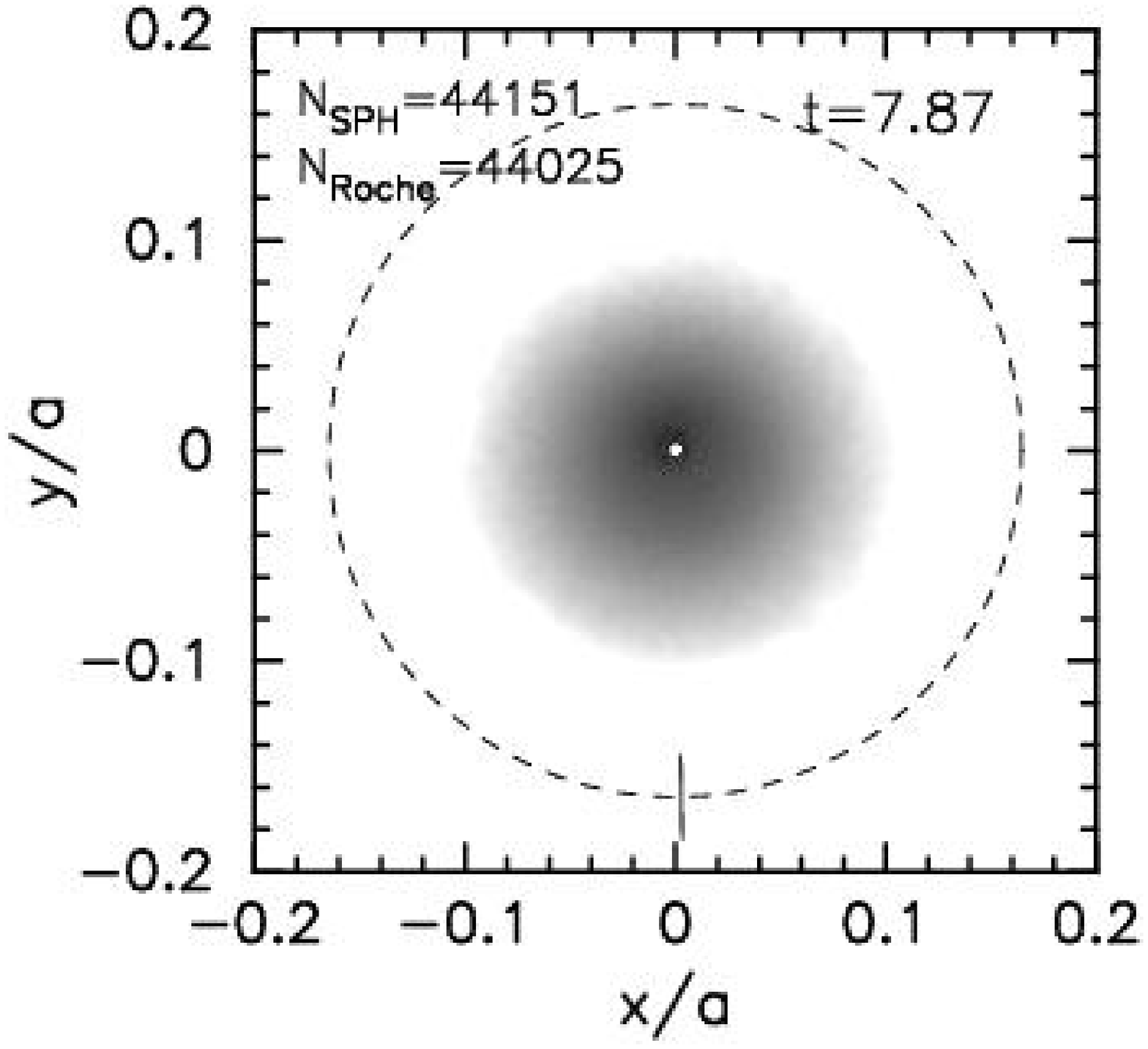}
}
 \caption{A sequence of snapshots of a developed accretion disc in
   model~1, which cover one orbital period ($t=7-8$). The format of
   the figure is the same as that of Fig.~\ref{fig:model6_ss}, except
   that the surface density is shown in a range of three orders of
   magnitude in the logarithmic scale.} 
 \label{fig:model1_ss}
\end{figure*}


\begin{figure*}
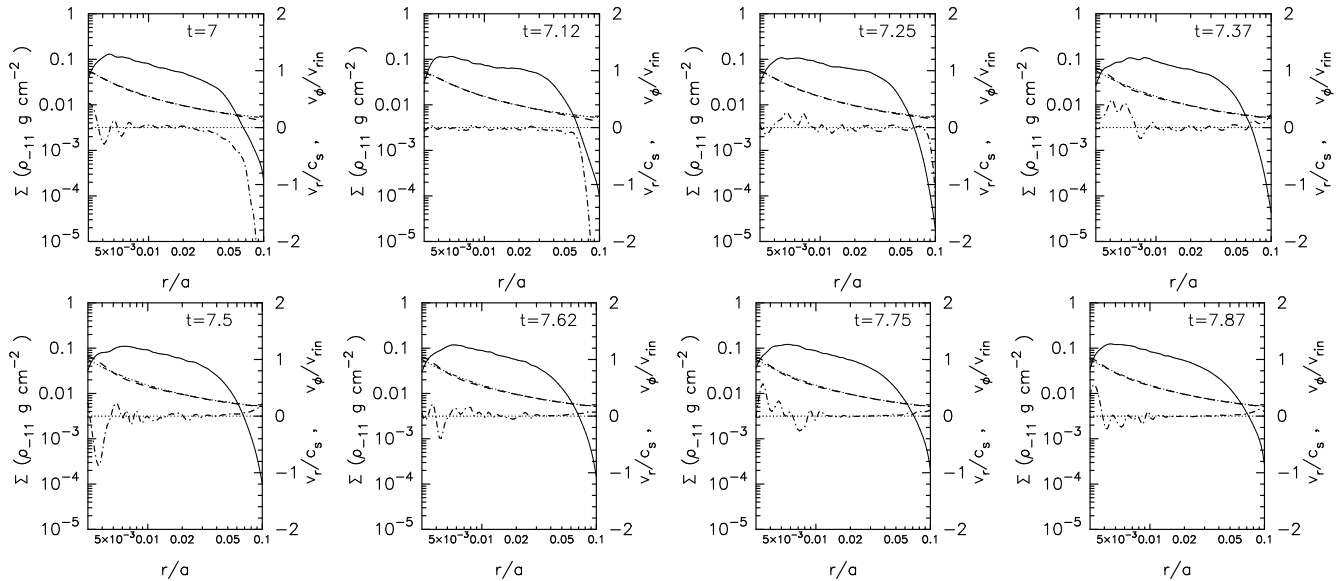

\resizebox{\hsize}{!}
{
\includegraphics*[angle=0,width=4.5cm,clip]{khfig29.ps}\hspace*{0.5em}
\includegraphics*[angle=0,width=4.5cm,clip]{khfig30.ps}\hspace*{0.5em}
\includegraphics*[angle=0,width=4.5cm,clip]{khfig31.ps}\hspace*{0.5em}
\includegraphics*[angle=0,width=4.5cm,clip]{khfig32.ps}}\\
\resizebox{\hsize}{!}
{
\includegraphics*[angle=0,width=4.5cm,clip]{khfig33.ps}\hspace*{0.5em}
\includegraphics*[angle=0,width=4.5cm,clip]{khfig34.ps}\hspace*{0.5em}
\includegraphics*[angle=0,width=4.5cm,clip]{khfig35.ps}\hspace*{0.5em}
\includegraphics*[angle=0,width=4.5cm,clip]{khfig36.ps}
}
 \caption{Time-dependence of the radial structure of the developed
  accretion disc shown in Fig.~\ref{fig:model1_ss}. The format of the figure is
  the same as that of Fig.~\ref{fig:model6_rad}.}
 \label{fig:model1_rad}
\end{figure*}

\subsection{Structure of a developed disc}
\label{sec:dev_disc}

Many Be/X-ray binaries with known orbital parameters have orbital 
eccentricities in the range of 0.3 to 0.9. In such systems, the 
mass-transfer rate from the Be disc is expected to have the strong phase
dependence \citep{oka2}. Accordingly, the structure of an
accretion disc formed around the neutron star in these systems is also
likely to be strongly phase dependent. In this section, we first 
discuss the phase-dependent structure of a developed accretion disc 
in model~1 and then compare that with those in other models.

\subsubsection{Model~1}
\label{sec:model1}

Fig.~\ref{fig:model1_ss} gives a sequence of snapshots of the accretion
flow around the neutron star covering one orbital period for $7 \le t
< 8$. The format of the figure is the same as that of
Fig.~\ref{fig:model6_ss}, except that the surface density is shown in the range 
of three orders of magnitude. We note from the figure that the disc shrinks 
at periastron passage by a negative torque exerted by the Be star and
restores its radius afterwards by the viscous diffusion. We also note
that the disc is significantly eccentric even after it is fully
developed, although the eccentricity is much smaller than that in the
initial phase of the disc formation.

Fig.~\ref{fig:model1_rad} shows the radial disc structure at the orbital
phases corresponding to Fig.~\ref{fig:model1_ss}. The format of the
figure is the same as that of Fig.~5. It is noted from
Fig.~\ref{fig:model1_rad} that the disc is nearly Keplerian at any phase
and the accretion flow is highly subsonic except in the outermost part
for a short period of time when the material transferred from the Be
disc falls towards the neutron star. We call this state of the
accretion disc developed. We have found that these features of a
developed disc hold for $t \ga 5$ in model~1. As in the initial phase
of the disc formation, the ram pressure by the supersonic infall of
matter transferred from the Be disc around periastron enhances the
density in the outermost part of the accretion disc, making the disc
outer edge sharp (see panels for $7.12 \le t \le 7.37$ in
Figs.~\ref{fig:model1_ss} and \ref{fig:model1_rad}). Afterwards, the disc
gradually expands until the next periastron passage of the Be star.

Fig.~\ref{fig:chardisc} shows how hot the disc is at $t=7.5$ in models~1 and 3-5. 
In each panel, the solid line, the dash-dotted line and the dashed line denote 
the disc temprature normalized by the initial temprature $T_{0}=1.3\times10^{4}\,\rmn{K}$, 
the ratio of the smoothing length to the harf-thickness of the disc $h/H$ 
and the relative disc thickness $H/r$, respectively.
From the upper-left panel of Fig.~\ref{fig:chardisc}, it is noted that 
the disc in model~1 is vertically resolved except for the innermost region of 
the disc ($r \la 5.0\times10^{-3}a$). 
The disc temprature varies from $\sim 10^{4}\,{\rm K}$ at $\sim 0.1a$ to 
$\sim 6.5\times10^{4}\,\rm{K}$ at the inner boundary. 
The disc is geometrically thin with the relative thickness of about 0.1-0.2.

\subsubsection{Models~2-5}
\label{sec:models2-5}

In this section, we describe the effects of the simulation parameters
in Table~\ref{tbl:models} on the structure of the accretion flow around
the neutron star. 

We ran model~2 to study the effect of the inner boundary on the disc
strcture. Fig.~\ref{fig:model2} shows the snapshots and the radial
structures of the accretion flow around the neutron star at $t = 7$
(periastron) and $t = 7.5$ (apastron) for model~2, which has the same
simulation parameters as model~1, except for a larger inner boundary of 
$r_{\rmn{in}} = 5.0\times10^{-3}$. Comparing Fig.~\ref{fig:model2}
with the panels at the corresponding times in
Figs.~\ref{fig:model1_ss} and \ref{fig:model1_rad} for model~1, we
note that the presence of the inner boundary artificially reduces the
density in a narrow region ($r \la 2 r_{\rmn{in}}$) near the boundary,
but outside of it, the disc structure is similar for both models.

\begin{figure}
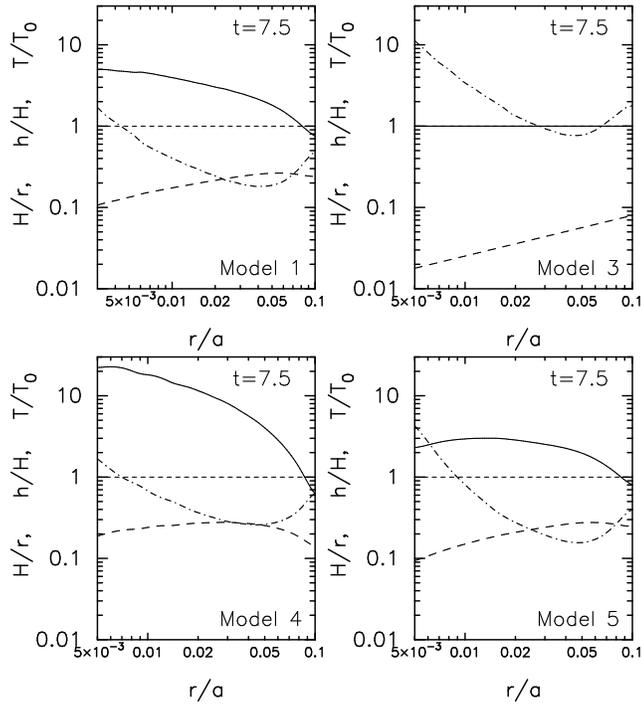

\resizebox{\hsize}{!}
{
\includegraphics*{rvkhfig1.ps}
\includegraphics*{rvkhfig2.ps}}\\
\resizebox{\hsize}{!}
{
\includegraphics*{rvkhfig3.ps}
\includegraphics*{rvkhfig4.ps}}
\caption{
Radial distributions of the disc thickness and temprature at $t=7.5$ in models~1 and 3-5. 
In each panel, the solid line, the dash-dotted line and the dashed line denote 
the disc temprature normalized by the initial temprature $T_{0}=1.3\times10^{4}\,\rmn{K}$, 
the ratio of the smoothing length to the harf-thickness of the disc $h/H$ 
and the relative disc thickness $H/r$, respectively.}
 \label{fig:chardisc}
\end{figure}

\begin{figure}
\resizebox{\hsize}{!}
{
\includegraphics*{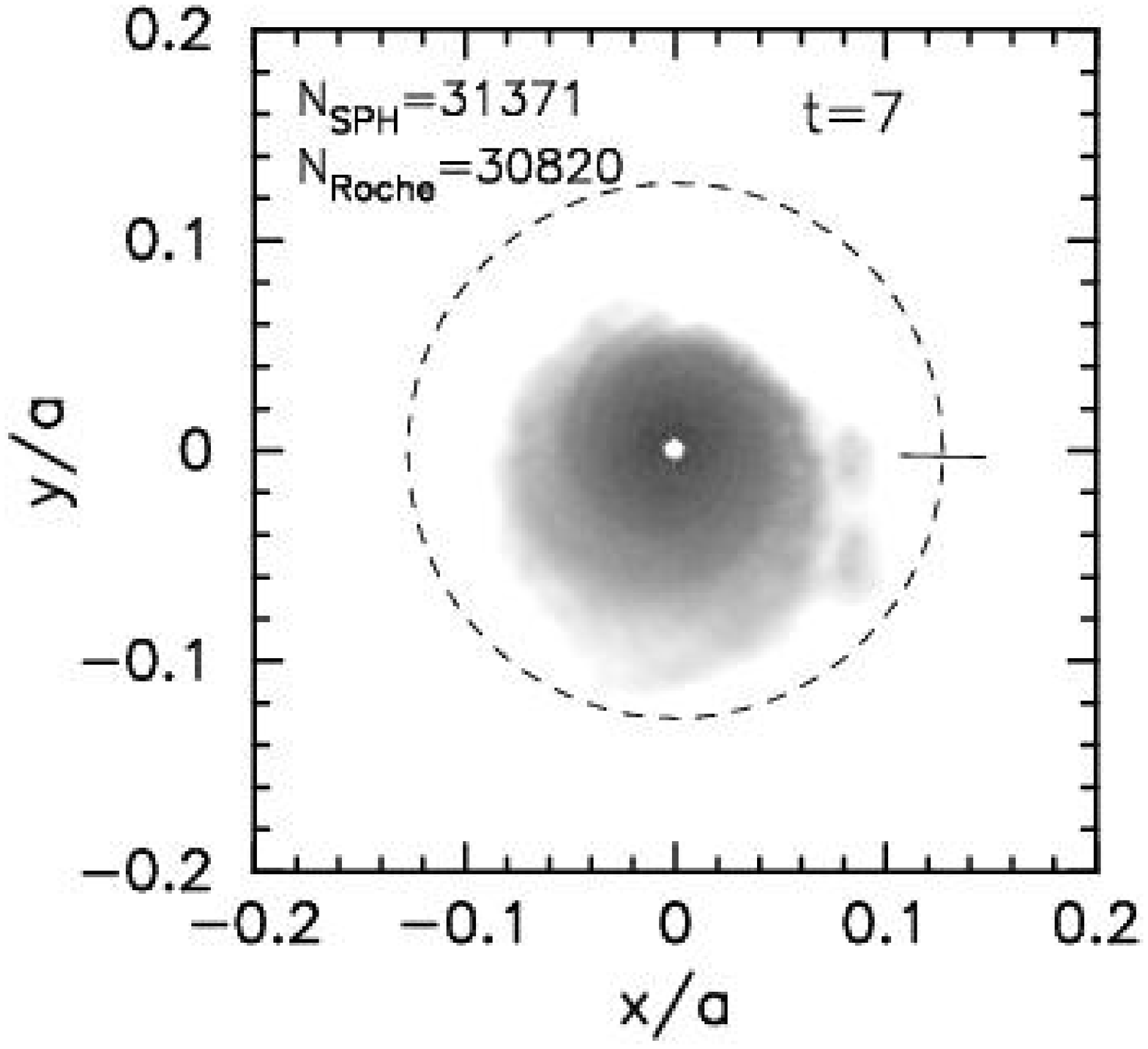}\hspace*{0.5em}
\includegraphics*{khfig38.ps}}\\
\resizebox{\hsize}{!}
{
\includegraphics*{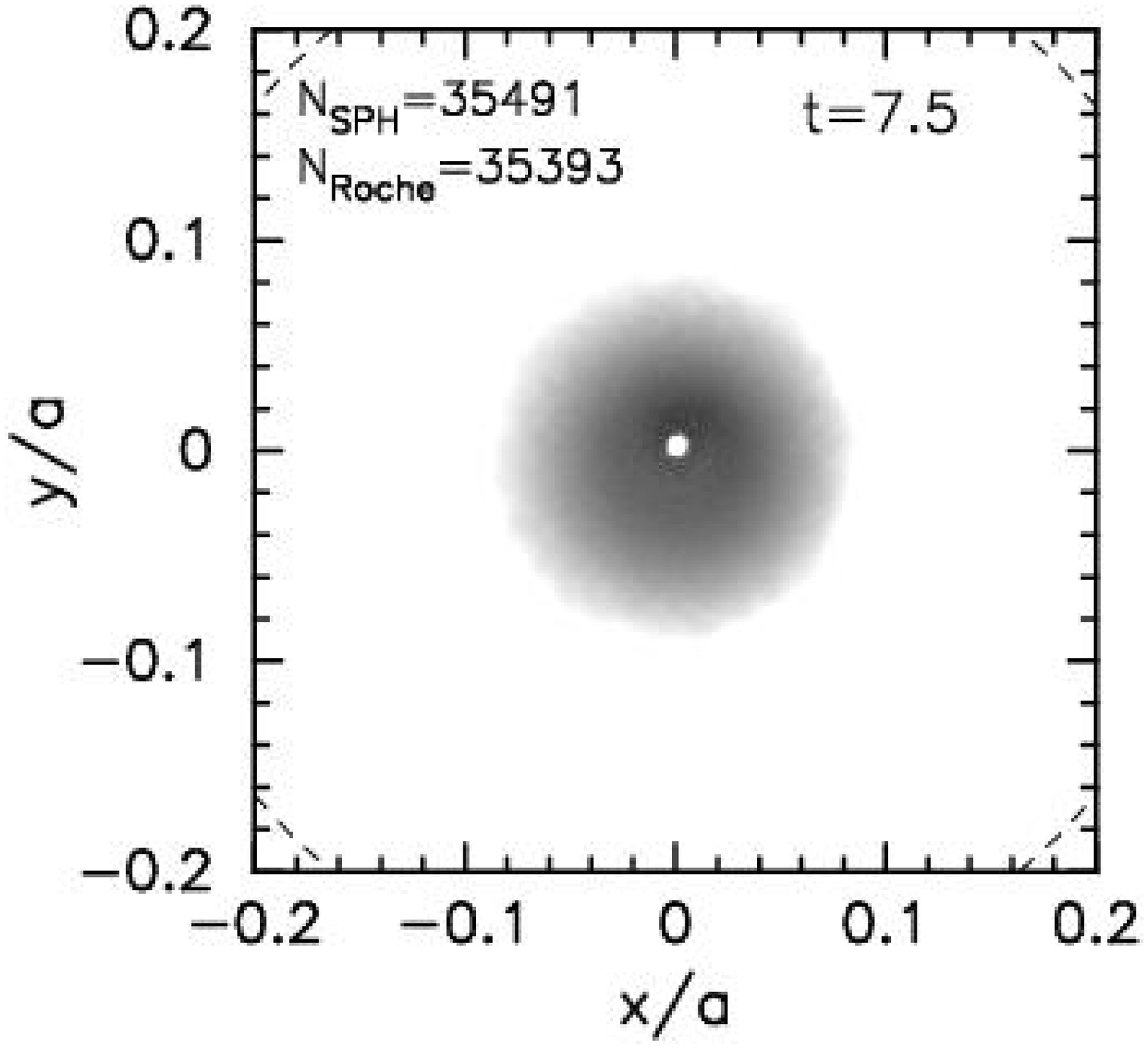}\hspace*{0.5em}
\includegraphics*{khfig40.ps}
}
 \caption{Time-dependence of the accretion disc structure in
  model~2. The top panels show the snapshot (left) and the radial
  structure (right) at the periastron, while the bottom panels show
  those at the apastron. In the snapshots, the surface density is
  shown in a range of three orders of magnitude in the logarithmic
  scale.}
 \label{fig:model2}
\end{figure}

\begin{figure}
\resizebox{\hsize}{!}
{
\includegraphics*{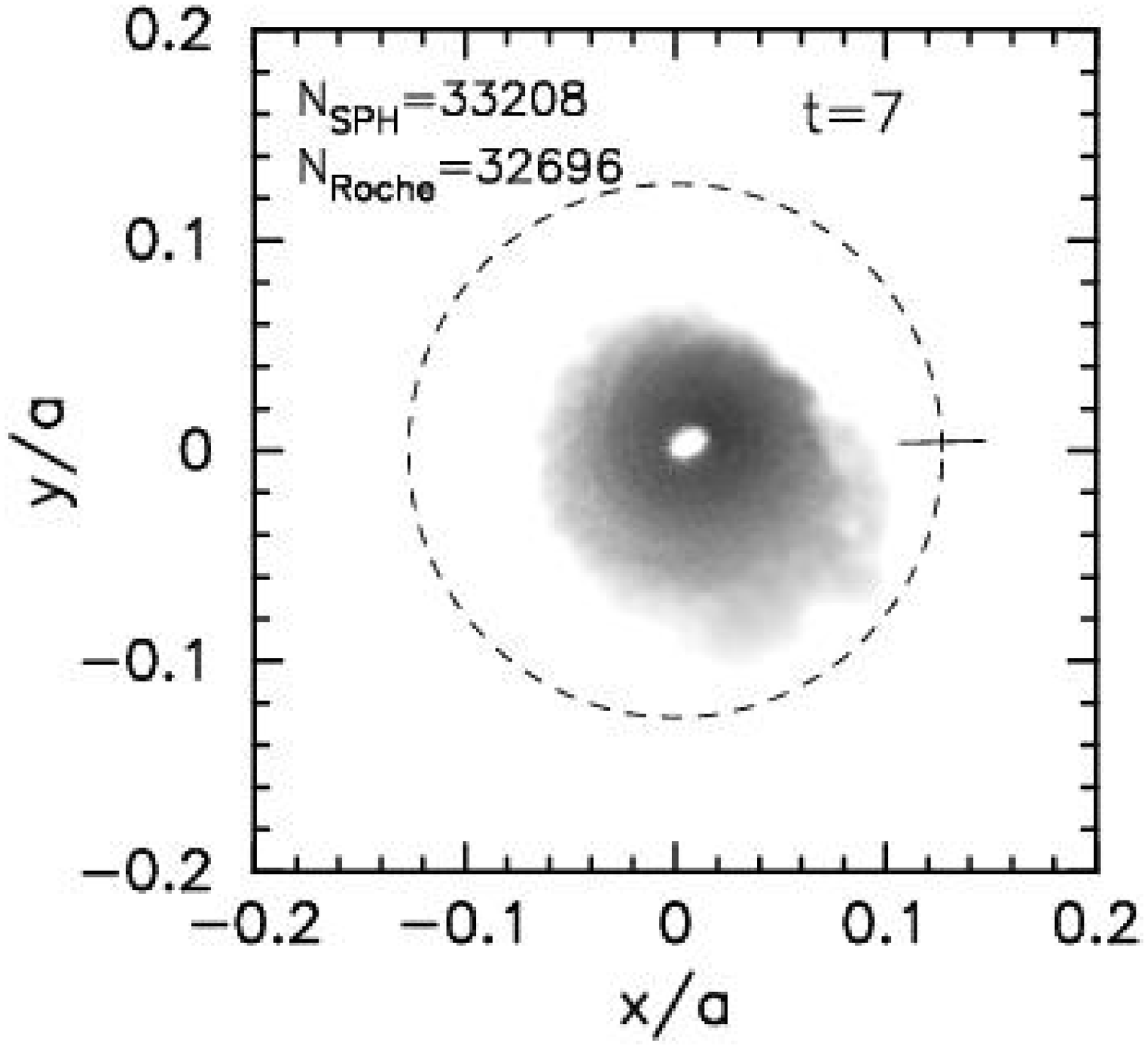}\hspace*{0.5em}
\includegraphics*{khfig42.ps}}\\
\resizebox{\hsize}{!}
{
\includegraphics*{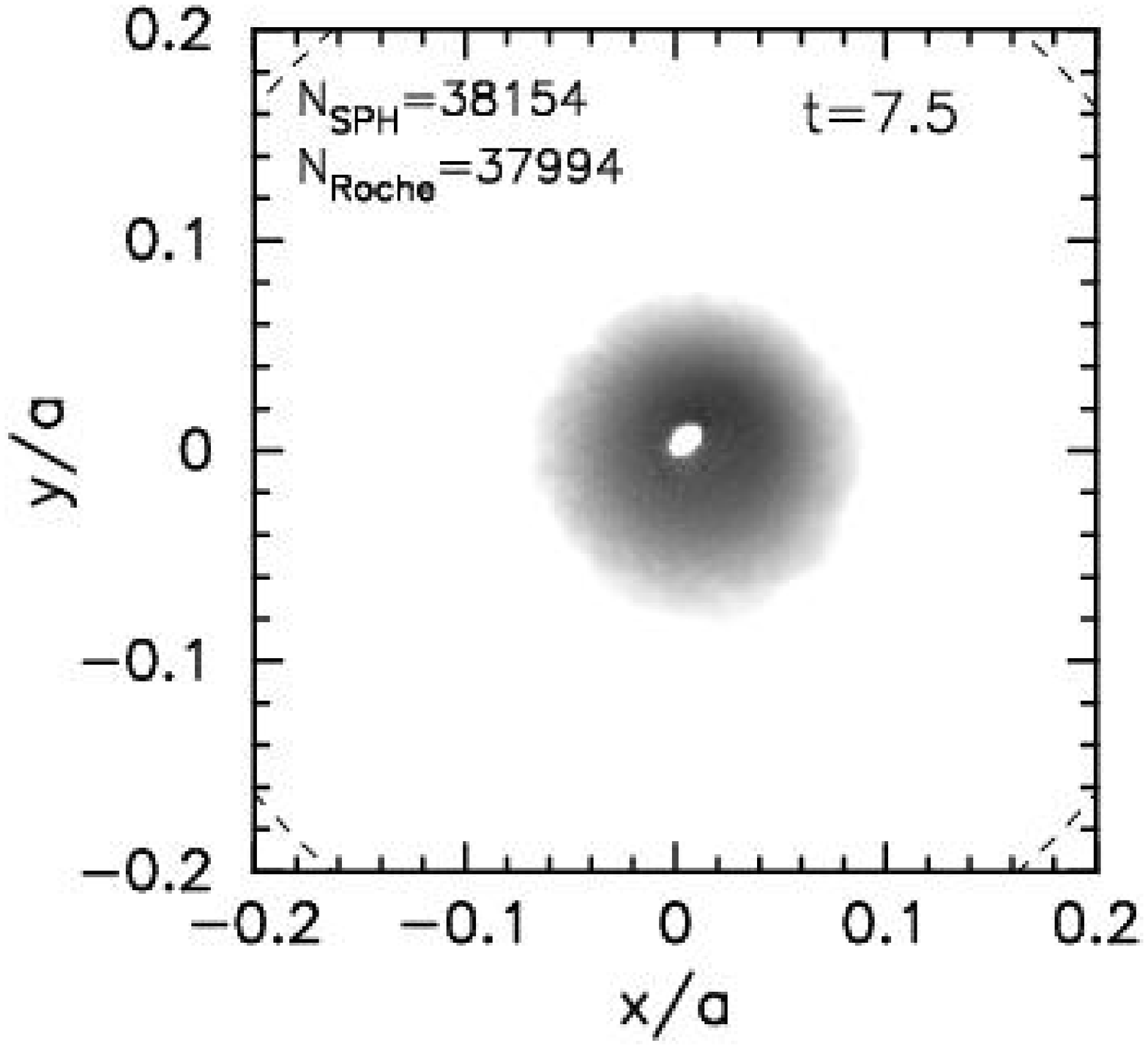}\hspace*{0.5em}
\includegraphics*{khfig44.ps}}
 \caption{Same as Fig.~\ref{fig:model2}, but for model~3.}
 \label{fig:model3}
\end{figure}

In order to study the dependence of the disc structure on the
polytropic exponent, we also ran models~3 ($\Gamma = 1$; isothermal
case) and 4 ($\Gamma = 5/3$; adiabatic case), for which the polytropic
exponent of 1.2 for models~1, 2 and 6 is in between. 
Fig.~\ref{fig:model3} shows the snapshots and the radial
structures of the accretion disc for model~3 (isothermal case) at the
same times as in Fig.~\ref{fig:model2}. From the figure, we note that
the distributions of the surface density and both of the azimuthal and
radial velocity components are similar to those in the initial phase
of the disc formation for $\Gamma = 1.2$ (see
Figs.~\ref{fig:model6_ss} and \ref{fig:model6_rad}). Consequently, in
model~3 the accretion disc is still developing at $t \sim 7$. This is
because the viscous time-scale of the isothermal disc, particularly in
an inner region, is longer than those in other models (see also
Fig.~\ref{fig:rcirc}). 
From the upper-right panel of Fig.~\ref{fig:chardisc}, 
it is noted that the disc in model~3 is not vertically resolved 
in most of the disc region. 

\begin{figure}
\resizebox{\hsize}{!}
{
\includegraphics*{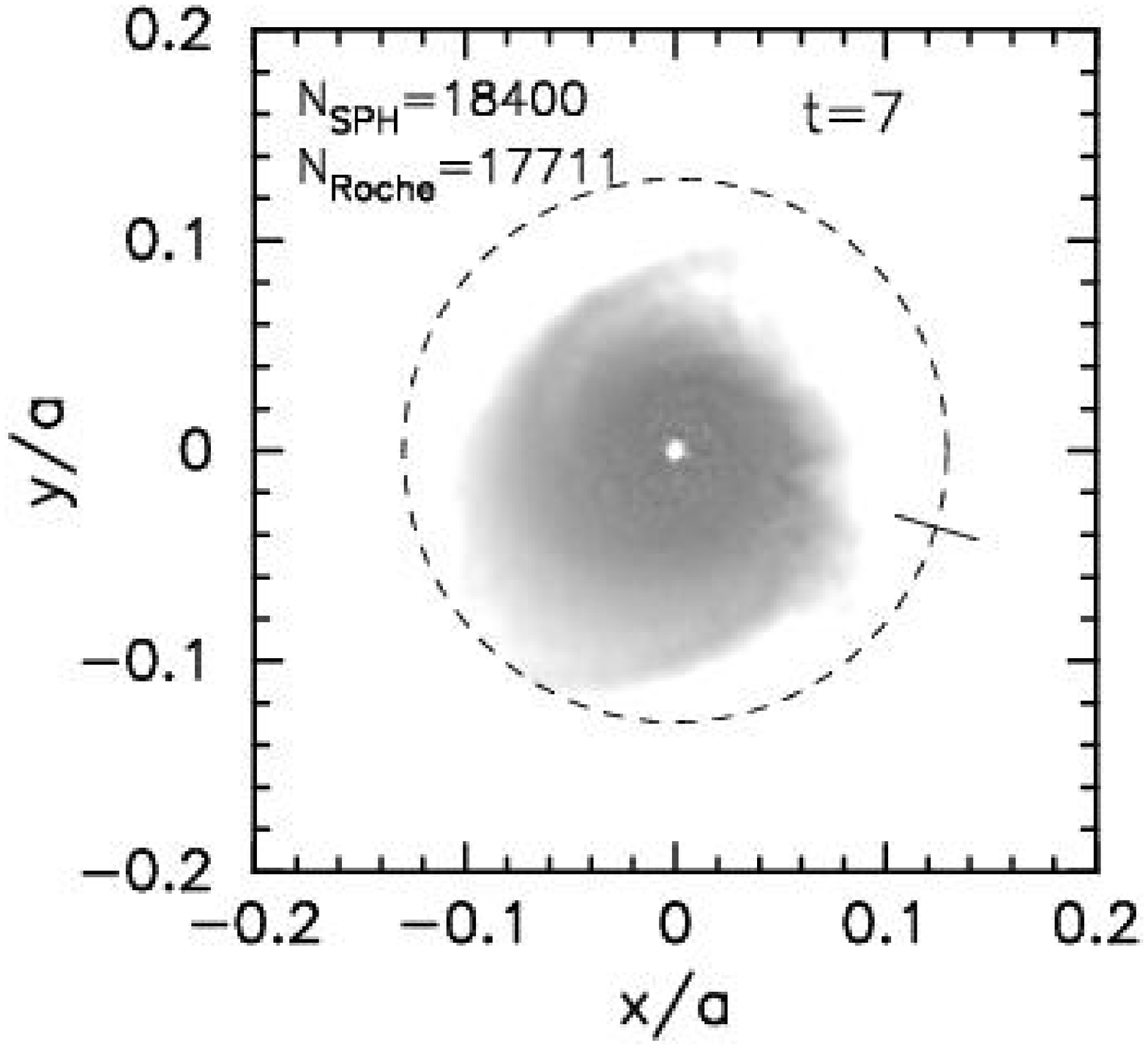}\hspace*{0.5em}
\includegraphics*{khfig46.ps}}\\
\resizebox{\hsize}{!}
{
\includegraphics*{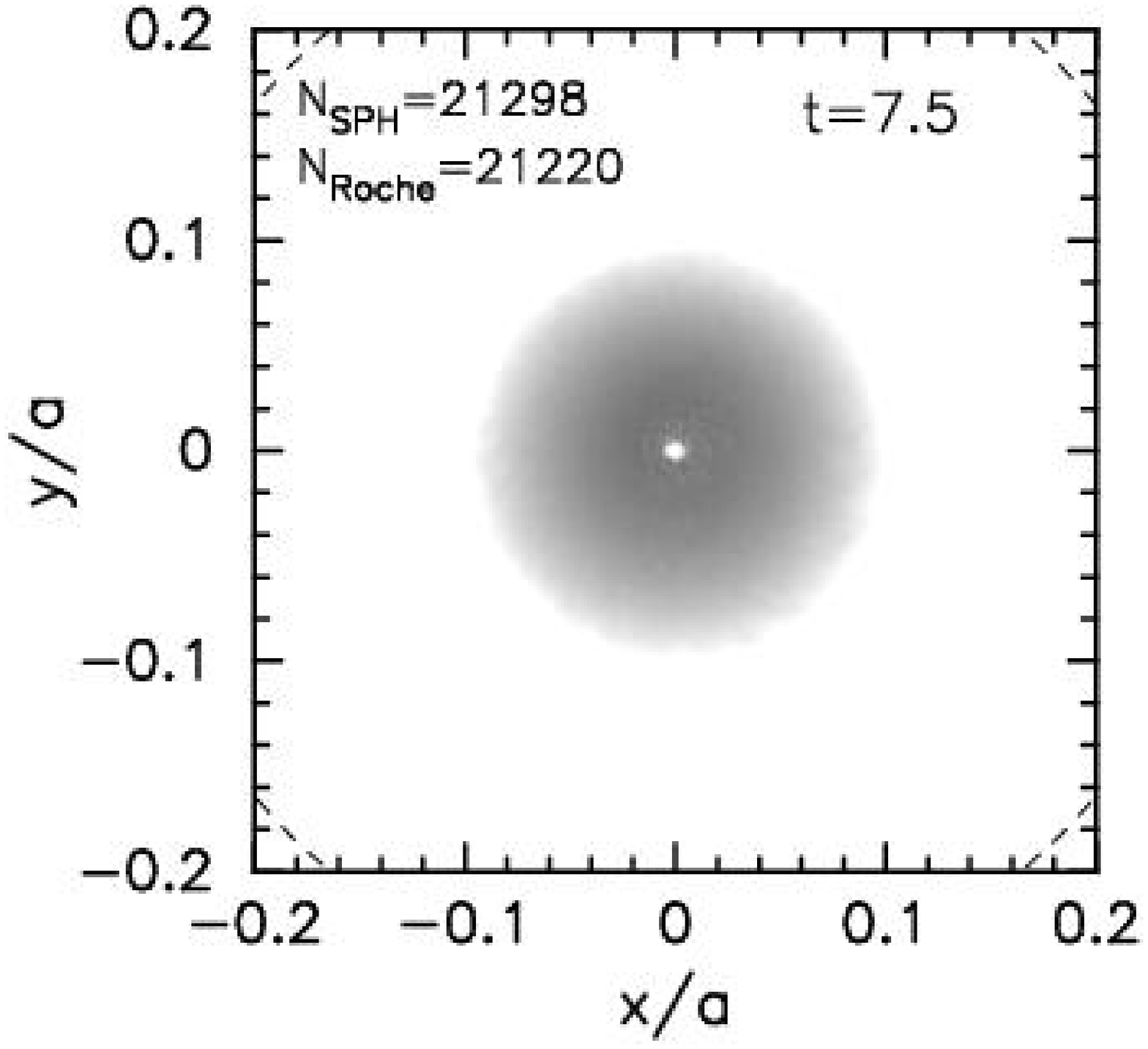}\hspace*{0.5em}
\includegraphics*{khfig48.ps}}
\caption{Same as Fig.~\ref{fig:model2}, but for model~4.}
 \label{fig:model4}
\end{figure}

Fig.~\ref{fig:model4} shows the disc structure for model~4 (adiabatic
case). We note that the disc is nearly Keplerian at both orbital
phases and that the surface density is lower and the disc is bigger
than those for model~2 with $\Gamma = 1.2$. This is due to a higher
viscous stress in model~4, resulting from a higher pressure. As
expected, in model~4, the accretion rate is much higher than the
models with lower polytropic exponents, which we will see in a later
section. 
From the lower-left panel of Fig.~\ref{fig:chardisc}, 
it is noted that 
the disc in model~4 is much hotter than in other models, reaching the temprature of 
$\sim5.0\times10^{5}\,\rm{K}$ near the inner boundary. 
The disc is vertically resolved except for the innermost region.

\begin{figure}
\resizebox{\hsize}{!}
{
\includegraphics*{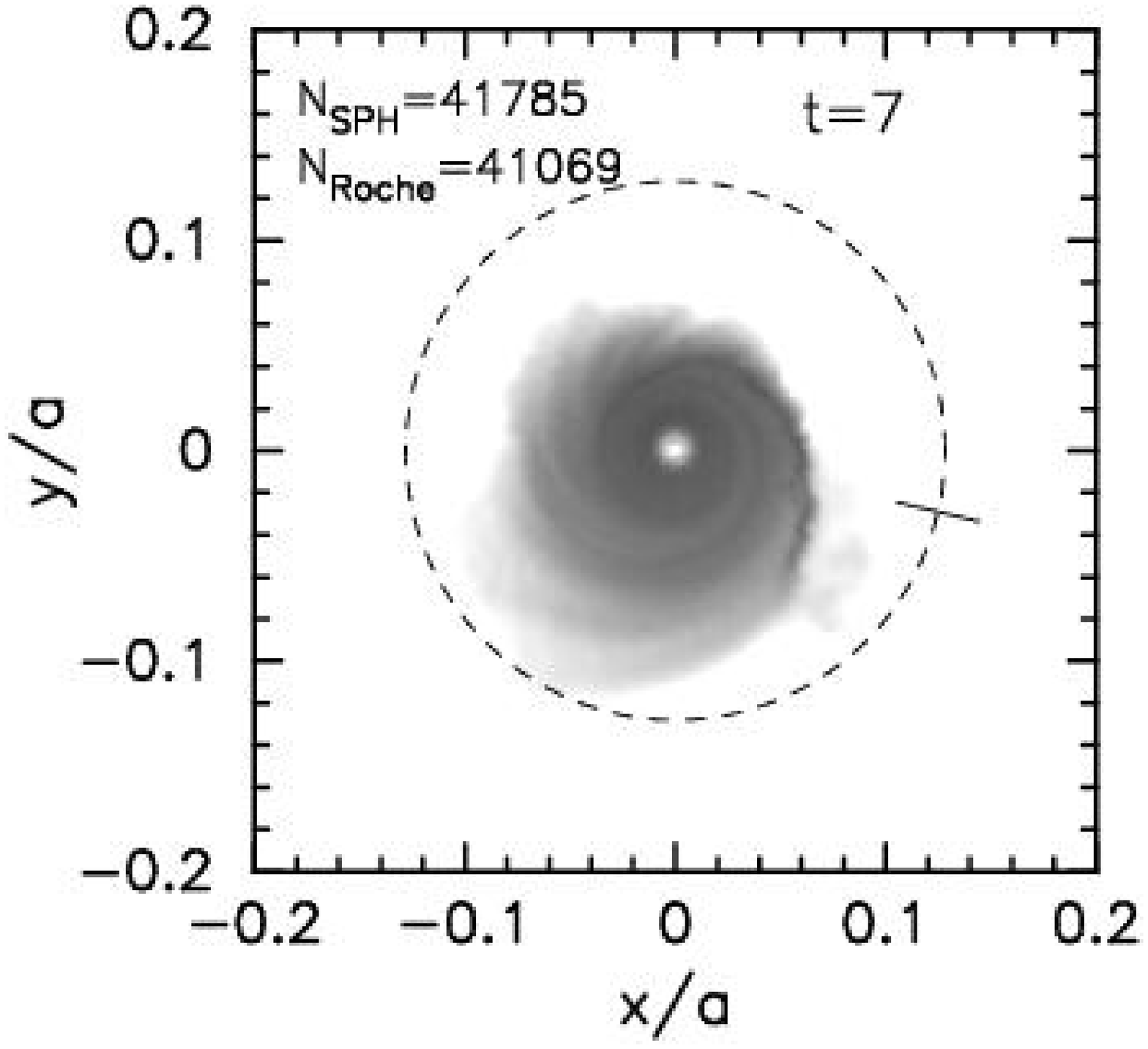}\hspace*{0.5em}
\includegraphics*{khfig50.ps}}\\
\resizebox{\hsize}{!}
{
\includegraphics*{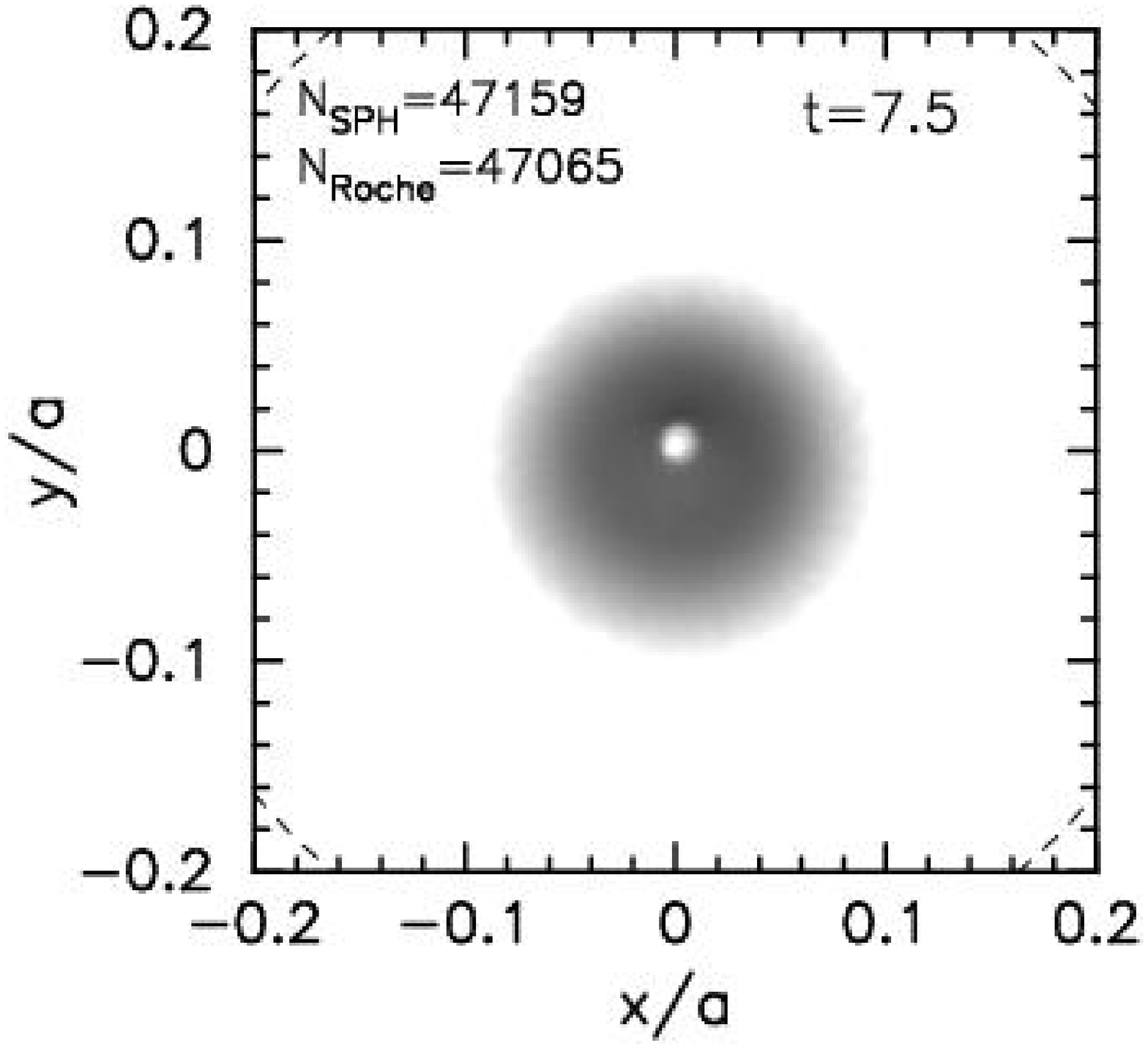}\hspace*{0.5em}
\includegraphics*{khfig52.ps}}
\caption{Same as Fig.~\ref{fig:model2}, but for model~5.}
 \label{fig:model5}
\end{figure}

We also studied a model with the standard, artificial viscosity
parameters. Model~5 has the same simulation parameters as those in
model~2, except that it has constant articficial viscosity parameters
$\alpha_{\rmn{SPH}} = 1$ and $\beta_{\rmn{SPH}} = 2$, for which
$\alpha_{\rmn{SS}}$ varies in time and space. Fig.~\ref{fig:model5}
shows the disc structure for model~5. 
From the figure and the lower-right panel of Fig.~\ref{fig:chardisc},  
it is immediately noted that the inner boundary affets the disc structure 
much more widely in this model. 
This is mainly because the inner disc region is not resolved
vertically for this number of SPH particles. Since the ratio of the
smoothing length to the harf-thickness of the disc $h/H$ in the inner region is 
much higher than unity, the shear viscosity parameter
$\alpha_{\rmn{SS}}$ in this model is much higher in the inner region
than that in other models.

Except for the above-mentioned dependence of the disc structure on the
simulation parameters, the accretion discs in our models share the
common features: They are nearly Keplerian after developed and their
structures modulate with the orbital phase. They are also
significantly eccentric. 
We will analyse the non-axisymmentry of the accretion discs 
in Be/X-ray binaries in a subsequent paper.


\subsubsection{Size of the accretion disc in Be/X-ray binaries}
\label{sec:adsize}

\citet{al} investigated the tidal/resonant truncation
of circumsteller and circumbinary discs in eccentric binaries and
found that a gap is always opened between the disc and the binary
orbit. Following their formulation, \citet{ne2} and 
\citet{oka1} have shown that the Be disc is truncated
via the resonant interaction with the neutron star, as long as
$\alpha_{\rmn{SS}} \ll 1$, which has recently been confirmed by
numerical simulations by \citet{oka2}.

It is interesting to study whether the tidal/resonant torque by the Be
star also truncates the accretion disc around the neutron
star. In order to have a measure of the accretion disc radius at which
the disc density has a major break, we have applied a non-linear least
squares fitting method to the radial distribution of the
azimuthally-averaged surface density $\Sigma$, adopting the following
simple fitting function,
\begin{equation}
   \Sigma \propto \frac{(r/r_{\rmn{d}})^{-p}}{1+(r/r_{\rmn{d}})^{q}},
   \label{eq:fitting}
\end{equation}
where $p$ and $q$ are constants and $r_{\rmn{d}}$ is the radius of
accretion disc. This method is the same as that adopted by \citet{oka2}.

\begin{figure}
\resizebox{\hsize}{!}{\includegraphics*{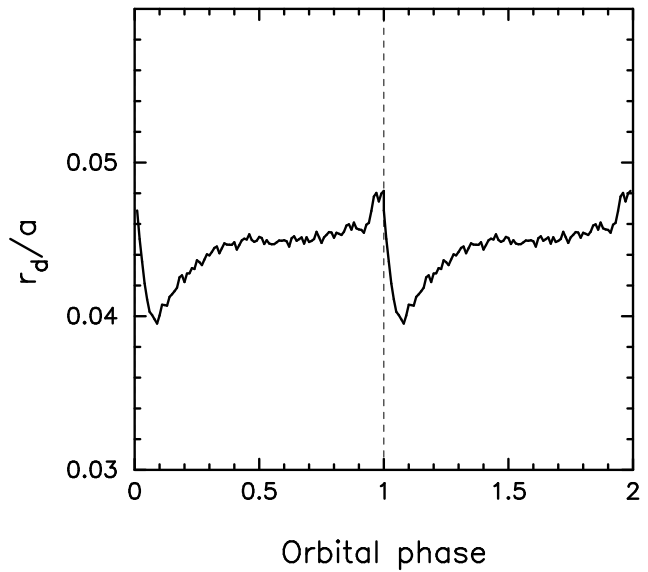}}
 \caption{Orbital-phase dependence of the accretion disc radius
            $r_{\rmn{d}}$. To reduce the fluctuation noise, the data
            was folded on the orbital period for $5 \le t \le 8$.}
 \label{fig:radius}
\end{figure}

Fig.~\ref{fig:radius} shows the orbital-phase dependence of the
accretion disc radius in model~1. To reduce the fluctuation noise, we
folded the data on the orbital period over $5 \le t \le 8$. It is
noted that the disc radius varies between $0.04 a$ and $0.05 a$. 
The disc shirinks at periastron due to a negative torque by the Be star 
and gradually restores its radius towards apastron 
due to the viscous diffusion. 
The rapid increase of the disc radius before periastron 
is due to the infall of matter from the Be star.

Next, we examine whether the disc radius shown in Fig.~13 
is determined by the resonant truncation 
as is the case in the Be disc in Be/X-ray binaries. 
Fig.~8 of \citet{al} gives the truncation radius of a circum-secondary disc as a
function of the orbital eccentricity $e$ and the disc Reynolds number
$Re$ for systems with the mass ratio of 0.1, which is similar to that
of Be/X-ray binaries. Since the accretion disc in our model is
geometrically thin and nearly Keplerian, the Reynolds number $Re$ is
given by $Re = (1/\alpha_{\rmn{SS}})(r/H)^2$ $\sim (1/\alpha_{\rmn{SS}})
(v_{\rmn{Kep}}/c_{\rmn{s}})^2$, where 
$v_{\rmn{Kep}}$ is the Keplerian velocity. In model~1, $Re
\sim 10^{3-3.5}$  for $0.01 a \la r \la 0.1 a$. From Fig.~8 of 
\citet{al}, with $e = 0.34$ and $Re \sim
10^{3-3.5}$, we have the truncation radius of $\sim 0.14 a$, which
corresponds to the 6:1 resonance radius. Note that this 
radius is much larger than the disc radius shown in
Fig.~\ref{fig:radius}. 
It is likely that the interaction with the Be star is significantly non-linear 
owing to its large mass. Then, the accretion disc would be truncated well inside 
the 6:1 radius given above. Even such a strong interaction, however,
 is unlikely to truncate the disc at a radius as small
 as that shown in Fig.~\ref{fig:radius}.

It is suggestive that the disc radius is about the same as the 
circularization radius $R_{\rm{circ}}$ of the matter transferred from
the Be disc after the periastron passage (see Fig.~\ref{fig:rcirc}).
Since the viscous time-scale at the circularization radius 
is much longer than the run time of our simulations, 
the matter once circularized dose not have time enough to change its radius 
significantly. 
Therefore, we conclude that the disc radius shown in Fig.~\ref{fig:radius} 
is determined by the specific angular momentum of 
the matter transferred from the Be disc.
It is probable, however, that the disc expands well 
beyond the circularization radius and is truncated by a tidal torque 
in a time-scale longer than the viscous one.


\begin{figure}
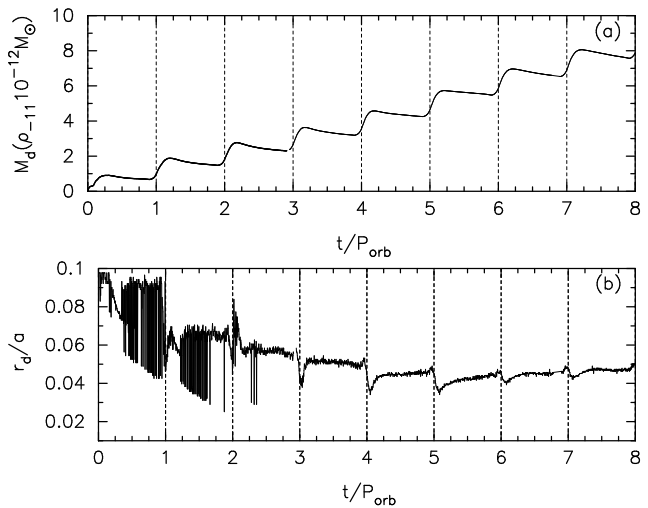

\resizebox{\hsize}{!}{\includegraphics*{khfig54.ps}}
\resizebox{\hsize}{!}{\includegraphics*{khfig55.ps}}
 \caption{Evolution of (a) the disc mass $M_{\rmn{d}}$ and (b) the
   disc size $r_{\rmn{d}}$ normalized by the semi-major axis for $0
   \le t \le 8$. In panel~(a), the disc mass is measured in units of
   $\rho_{-11}10^{-12}M_{\odot}$.}
 \label{fig:masrad}
\end{figure}

\subsection{Disc evolution}
\label{sec:evolv}

As mentioned in Section~1, most of the Be/X-ray binaries show
transient X-ray activities, of which nature is considered to result
from the interactions between the accreted material and the rotating
magnetized neutron star \citep{st}. In this section, we
first describe the growth of the accretion disc in mass and radius. We
then examine the evolution of the mass-accretion rate obtained by our
simulations, from which we finally discuss the X-ray activity
expected in Be/X-ray binaries.

\subsubsection{Growth of the disc}
\label{sec:growth}

Fig.~\ref{fig:masrad}(a) shows the evolution of the disk mass in
model~1 over the period of $0 \le t \le 8$. In the long term, the disc
mass increases with time, because the the gas transferred from the Be
star accumulates in the accretion disc owing to the viscous 
time-scale longer than the orbital period. On the other hand, in the
short term, it modulates with the orbital phase. The disk mass starts
increasing, when the mass-supply from the Be disc begins slightly
before every periastron passage. It shows a rapid increase for a short
while, followed by a gradual decrease by accretion with no mass-supply
from the Be disc, which lasts until the next periastron passage. 
At first glance, the disc mass in model~1 seems to increase secularly and 
approach no steady average state. 
On closer inspection, however, we note that the increasing rate 
of the disc mass gradually decrases with time after the disc is developed. 
This suggests an interesting possibility that the disc reachs a steady average state 
in a time-scale much longer than the term shown in Fig.~14(a).

In Fig.~\ref{fig:masrad}(b), we show the evolution of the disc radius
over $0 \le t \le 8$. One may expect that the disc radius gradually
increases with the disc mass. From the figure, however, we note that
this is the case only for $t \ga 5$. In fact, for $t \la 5$, the disc
radius decreases as the disc grows. This counter-intuitive
decrease in radius is related to the change in the disc
eccentricity. In the early stage of disc formation, the disc/ring is
highly eccentric (see Fig.~\ref{fig:model6_ss}), for which the radius
$r_{\rmn{d}}$ calculated by equation~(\ref{eq:fitting}) is larger than
that for a circular disc with the same disc mass. Since the 
disc eccentricity decreases with time, so does the disk radius. In
model~1, the effect of the disc circularization is balanced with that
of the disc growth at $t \sim 5$. After this epoch, the disc radius
increases with the disc mass. Thus, the evolution of the accretion
flow around the neutron star in Be/X-ray binaries involves a two-stage
process, which consists of the initial developing stage and the later
developed stage.

\subsubsection{Phase-dependent accretion}
\label{sec:accretion}

Fig.~\ref{fig:mdot_acc}(a) shows the evolution of the mass-accretion
rate and the corresponding X-ray luminosity in model~1 for $0 \le t
\le 8$. Here we calculated the X-ray luminosity by 
$L_{\rm{X}}=\eta GM_{\rm{X}}\dot{M}_{acc}/R_{\rm{X}}$ 
with the X-ray emission effciency $\eta=1$ for the neutron star.

The thick and thin lines denote the mass-accretion rate in
models~1 and 2, respectively. Note that the mass-accretion rate in our
simulations generally has double peaks per orbit. The first peak,
which occurs at the periastron, is due to the contraction of the disc
by the tidal torque of the Be star, except that the peaks at
periastron during the initial phase of disc formation are mainly due
to the direct accretion of gas with low specific angular momentum. The
second peak, which lags behind the peak in the mass-transfer rate from
the Be disc shown in Fig.~\ref{fig:mdot_cap}, results from the viscous
accretion of matter transferred from the Be disc. The first peak could
be artificial since it is related to the presence of the inner
simulation boundary.

The orbital modulation in the mass-accretion rate is significantly
different between the developing disc and the developed disc. The
first peak becomes weak as the disc developes. After the disc is fully
developed, the second peak is higher than the first peak.

Fig.~\ref{fig:mdot_acc}(a) also shows that, once developed, the
accretion disc in 4U\,0115+63 shows the peak X-ray luminosity of $L_X
\la 2 \cdot 10^{35}\,\rmn{erg\,s}^{-1}$. This level of X-ray emission
enters the transition regime (see \citealt{ca}) and is
consistent with the observed X-ray luminosity in the quiescent state.

For comparison purpose, we present in Fig.~\ref{fig:mdot_234} the
evolution of the mass-accretion rate and the corresponding X-ray
luminosity in models~2-4 for $0 \le t \le 8$. The thick and thin solid
lines and the thin dotted line are for models~2, 3 and 4,
respectively. We see from the figure that the accretion rate 
increases with increasing polytropic exponent. In models~2 ($\Gamma =
1.2$) and 3 ($\Gamma = 1$), the accretion rate profile has the double
peaks as described above. The long-term change in the peak profiles is
also similar to that in model~1. We note that the disc in model~2
entered the stage of the developed disc at $t \sim 6$. On the other
hand, the disc in model~3 is at the developing stage even at $t = 8$,
which is consistent with the disc structure shown in
Fig.~\ref{fig:model3}.

In contrast to the models with low $\Gamma$, the accretin rate in
model~4 has a single and periodic peak per orbit, even after the disc is
developed. With $\Gamma=5/3$, the pressure in the disc in model~4 is
much higher than that in other models. This means that much larger
viscous torque exerts on the disc in model~4 than in other models,
because the viscous stress in the $\alpha$-viscosity prescription is
proportional to the pressure. As a result, the viscous torque
dominates the tidal torque in model~4 even at periastron. This is
why there is no peak at periastron and the accretion rate shows
the regular orbital modulation with the amplitude much higher than in 
other models.

It is important to note that the disc evolution and its orbital
modulation is sensitive to the polytropic exponent $\Gamma$. This
strongly implies that the cooling effciency in the accretion disc
plays an important role in the X-ray activity in Be/X-ray binaries.

\subsubsection{Magnetospheric radius}
\label{sec:magneto}

Changes in the mass flux on the magnetosphere of the neutron
star determine whether the system is in the direct accretion regime or
in the propeller regime. While the matter falls directly on to the
neutron star in the direct accretion regime, no material accretes on
to the neutron star in the propeller regime by the centrifugal
inhibitation. If the magnetospheric radius $R_{\rmn{M}}$ is smaller
than the corotation radius $R_{\Omega}$, the system is in the
direct accretion regime, otherwise, in the propeller regime.

Assuming the neutron star magnetic field as the dipole-like field, the
magnetospheric radius $R_{\rmn{M}}$, obtained by equating the magnetic
field pressure of the neutron star to the ram pressure of the
accreting matter from the accretion disc, is written by
\begin{eqnarray}
   R_{\rmn{M}} 
   &=&
   5.1 \times 10^{8} k \left( \frac{\dot{M}}{10^{16}\,\rmn{erg\,s^{-1}}}
   \right)^{-2/7}
   \left( \frac{M_{\rmn{X}}}{M_{\odot}} \right)^{-1/7} \nonumber \\ 
   &\times& 
   \left(\frac{\mu}{10^{30}\,\rmn{G\,cm}} \right)^{4/7}
   \hspace{1mm}\ \rmn{cm},
   \label{eq:magnetosphere}
\end{eqnarray}
(\citealt{pr}; \citealt*{la}; see also \citealt{fr}), 
where $\mu$ is the magnetic moment. Here, $k$ is a
constant model parameter of order unity, depending on the physics and geometry of accretion. 
For spherical accretion, $k \simeq 0.91$. The original model
calculations by \citet{gh} for disc accretion on to the
aligned rotating magnetized neutron star obtained $k \simeq 0.47$,
whereas calculations by \citet{os}, which included an
induced wind not considered by \citet{gh}, resulted in the value of
$k \simeq 0.923$ for disc accretion.

\begin{figure*}
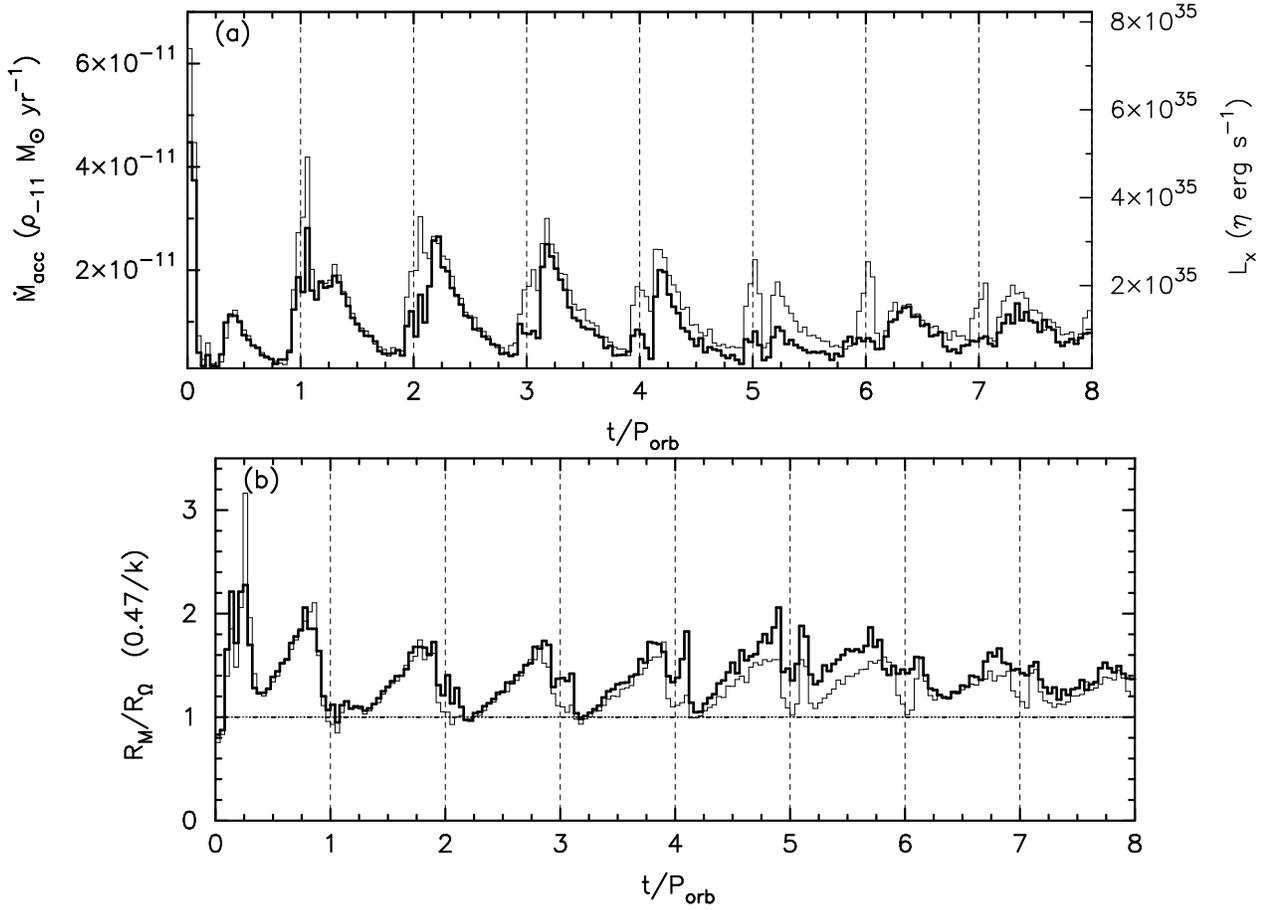

\centerline{\includegraphics*[height=6.0cm,clip]{khfig56.ps}}
\centerline{\includegraphics*[height=6.0cm,clip]{khfig57.ps}}
 \caption{Evolution of (a) the mass accretion rate
   $\dot{M}_{\rmn{acc}}$ and (b) the ratio of the magnetosphric
   radius to the corotation radius $\rm{R_{M}}/\rm{R_{\Omega}}$. 
   In each panel, the thick and thin lines are for models~1 and 2,
   respectively. In Panel~(a), the mass-accretion rate is measured in
   units of $\rho_{-11} M_{\odot}\,\rmn{yr}^{-1}$. The right axis is
   for the X-ray luminocity corresponding to the mass-accretion rate
   with $\eta=1$. In Panel~(b), the horizontal dotted line denotes the
   critical line of $R_{\rm{M}}=\rm{R_{\Omega}}$.}
 \label{fig:mdot_acc}
\end{figure*}

Fig.~\ref{fig:mdot_acc}(b) shows the evolution of the magnetospheric
radius $R_{\rmn{M}}$ normalized by the corotation radius $R_{\Omega}$,
where $k = 0.47$ was adopted. In the figure, the thick and thin lines
are for models~1 and 2, respectively. Recall that these models have
the same parameters except for the radius of the inner simulation
boundary $r_{\rmn{in}}$; $r_{\rmn{in}} = 3 \times 10^{-3}$ in model~1, 
while $r_{\rmn{in}} = 5 \times 10^{-3}$ in model~2. It should be noted
that the magnetospheric radii in both models evolve in a very similar
way, despite the different inner radii. Our result on the
magnetospheric radius is robust in this sense.

\begin{figure*}
\resizebox{\hsize}{!}{\includegraphics*{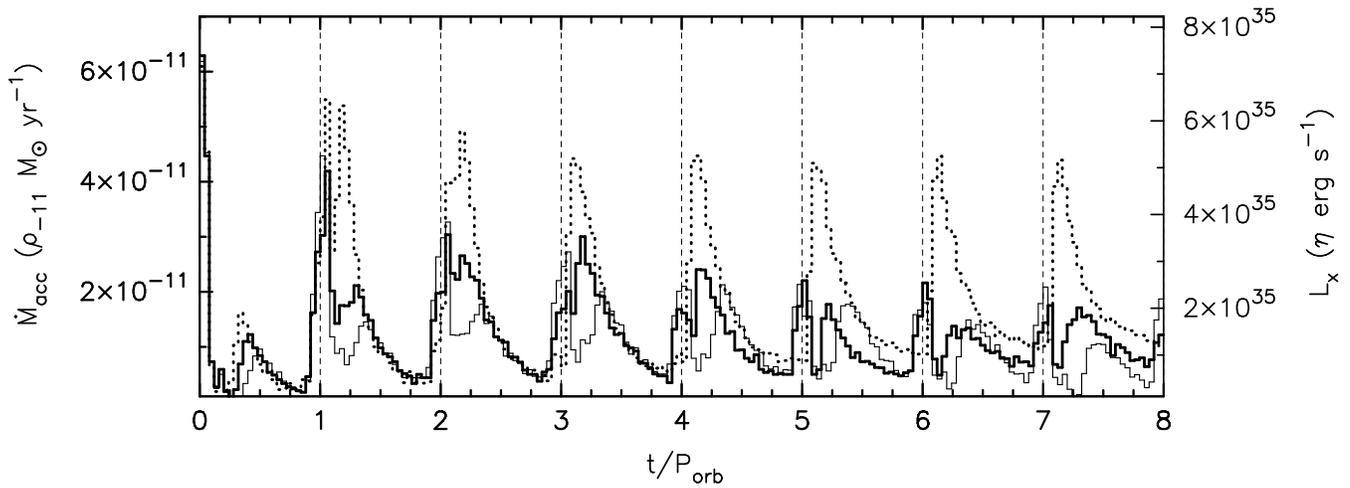}}
 \caption{Evolution of the mass-accretion rate $\dot{M}_{\rmn{acc}}$
   for models~2-4 with different polytropic exponents. The thick and
   thin solid lines and the thick dotted line denote the
   mass-accretion rates for models~2, 3 and 4, respectively. The
   mass-accretion rate is measured in units of $\rho_{-11}
   M_{\odot}\,\rmn{yr}^{-1}$. The right axis is for the X-ray
   luminocity corresponding to the mass-accretion rate with $\eta=1$.}
 \label{fig:mdot_234}
\end{figure*}

We also note from Fig.~\ref{fig:mdot_acc}(b) that the ratio of the
magnetospheric radius to the corotation radius,
$R_{\rmn{M}}/R_{\Omega}$, varies between $\sim 2k$ and $\la
4k$. Consequently, for a plausible range of $k$ ($0.5 \la k \la 1$),
the magnetospheric radius in models with $\Gamma \le 1.2$ is almost
always larger than the corotation radius so that the accretion on to
the neutron star is inhibited, which is consistent with the oberved 
X-ray luminosity of 4U\,0115+63 in quiescence (\citealt{ca}). On the
other hand, we can see from equation~(\ref{eq:magnetosphere}) and
Fig.~\ref{fig:mdot_234} that model~4 with $\Gamma = 5/3$ is consistent
with the observed X-ray behaviour of 4U\,0115+63 only if $k \ga
0.6$. For $k \sim 0.5$, the neutron star in model~4 enters the direct
accretion regime after every peristron passage of the Be star. Then,
the system would show regular, periodic X-ray outbursts at a level of
$\sim 5 \times 10^{35} \rmn{erg\,s}^{-1}$, which has not been observed
for 4U\,0115+63.


\section{summary and discussion}
\label{sec:summary}

For the first time, we have performed numerical simulations of 
accretion on to the neutron star in Be/X-ray binaries, taking into
account the phase-dependent mass transfer from the Be disc. We have
adopted the mass-transfer rate from a high-resolution simulation by 
\citet{oka2} for a coplanar system with a short period
($P_{\rmn{orb}}=24.3\,\rmn{d}$) and moderate eccentricity $(e=0.34)$,
which targeted 4U\,0115+63, one of the best studied Be/X-ray
binaries. We have found that a nearly Keplerian, non-steady accretion
disc is formed around the neutron star regardless of the parameters
adopted. The formation of the accretion disc involves a two-stage
process, which consists of the initial developing stage and the later
developed stage. The disc structure shows a strong dependence on the 
orbital phase, because both the tidal potential and the mass-transfer
rate change periodically. The disc is also significantly eccentric, 
because the gas particles from the Be disc originally have elliptical 
orbits.

In 4U\,0115+63, Type~I X-ray outbursts have occurred only as a short
series after occasional Type~II X-ray outbursts \citep{ne1}.
Except for such occasions, the system is in the quiescent state 
($10^{33} - 10^{35}\ \rmn{erg\,s}^{-1}$) \citep{ca}. 
The peak X-ray luminosity estimated from our simulations corresponds to 
the transition regime between the direct accretion regime and the 
propeller regime  and is consistent with the observed X-ray behaviour.
Note, however, that the discussion based on the X-ray luminosity
estimated from the simulations only gives the necessary condition for
the quiescent state.

In order to see whether our model satisfies a more stringent
condition, we also have estimated the magnetospheric radius by using a
general formula with the mass-accretion rate from our simulations and
compared with the corotation radius for 4U\,0115+63. 
The result was that our models with $\Gamma \le 1.2$ are consistent
with the observed X-ray behaviour.  Other models with $\Gamma = 5/3$
and $k \la 0.6$ were ruled out, because in these models the system
would show an X-ray outburst after every periastron passage, which has
never been observed.
We have to admit, however, that the radius of our inner
simulation boundary is about 60-100 times as large as the corotation
radius of the neutron star in 4U\,0115+63. To make the constraint on
the system parameters more reliable, we need numerical simulations
with a much smaller inner boundary.

It is important to note that our model is not only consistent with the
quiescent state of 4U\,0115+63. It can also give a natural explanation
for a short series of Type~I X-ray outbursts after occasional Type~II X-ray 
outbursts. The mass-transfer rate from the Be disc in the X-ray
outbursts is likely much higher than in the quiescent state. Suppose
that the mass-transfer rate from the Be disc in 4U\,0115+63 is
temporarily increased by an order of magnitude. Then, the estimated
X-ray luminosity in our model increases by an order of magnitude to a
level of several $\times 10^{36}\,\rmn{erg\,s}^{-1}$, a typical X-ray
luminosity of Type~I X-ray outbursts. At the same time, the magnetosphric
radius $R_{\rmn{M}}$ decreases by a factor of about two, for which
$R_{\rmn{M}}$ is almost always smaller than the corotation
radius, $R_{\Omega}$, if $k \sim 0.5$ (see
Fig.~\ref{fig:mdot_acc}(b)). This means that the system is in the
direct accretion regime. Thus, our model explains the Type~I X-ray outbursts
in 4U\,0115+63 as a temporal phenomena with an enhanced mass-transfer
from the Be disc.

Be/X-ray binaries are distributed over a wide range of orbital periods 
$(16\,{\rmn{d}} \la P_{\rm{orb}} \la 243\,{\rmn{d}})$ and eccentricities
$(e \la 0.9)$. It is interesting to consider here the effects of
orbital parameters. In this paper, we have seen that the main
ingredients which determines the X-ray behaviour of a system are the
viscous time-scale relative to the orbital period and the mass-transfer
rate from the Be disc. Since the former is insensitive to the orbital
period itself (see equation~(\ref{eq:tratio})), the latter is most
important if the viscosity parameter and the equation of state are
similar in accretion discs of Be/X-ray binaries.  

In systems with $\alpha_{\rmn{SS}} \la 0.1$, the viscous time-scale in
an outer part of the accretion disc is much longer than the orbital
period unless the cooling is ineffective and the gas obeys an
adiabatic equation of state. In these systems, persistent accretion
discs are expected to form. Then, the mass-transfer rate from the Be
disc determines whether the system shows Type~I X-ray outbursts. For
example, for a system with a moderate orbital eccentricity of $e =
0.34$ and a typical density of Be disc, the peak mass-transfer rate is
$\sim 2 \times 10^{-10} M_{\sun}\,{\rmn{yr}}^{-1}$, for which the
system stays in quiescence. It is obvious, however, that the situation
should be different in systems with much higher rate of
mass-transfer. A high mass-transfer rate is expected to result from a
high orbital eccentricity and/or an intrinsically dense Be
disc. Therefore, it is not strange that many systems which have
regularly shown Type~I X-ray outbursts have orbital eccentricities higher
than that of 4U\,0115+63 studied in this paper. In a subsequent paper,
we will perform numerical simulations of accretion in systems with
high orbital eccentricities. 

Above we have assumed that the viscous time-scale is much longer than 
the orbital period in all Be/X-ray binaries. However, it is possible
that an accretion disc has a relatively short viscous time-scale by
some reason. One possibility is that the disc has a low cooling
efficiency and becomes hot, particularly in an inner region. 
Our model~4 with $\Gamma = 5/3$ gives an extreme example of 
such discs (see also Fig.~\ref{fig:chardisc}).
As shown in Fig.~\ref{fig:mdot_234}, in a system with a short
viscous time-scale, the mass-accretion rate has a high and sharp peak
shortly after every periastron passage. The resemblance between the
accretion rate profile in model~4 and the X-ray light curves of
regular Type~I X-ray outbursts from systems such as EXO\,2030+375 \citep{wi1} 
implies that these are systems with relatively short
viscous time-scales.

Be/X-ray binaries are an ideal group of objects for studying the
physics of accretion under the circumstances which have been studied
little. Unlike close binaries, many Be/X-ray binaries have eccentric
orbits. They can also be highly inclined systems. Such systems provide
us a valuable oppotunity to study the effects of the
periodically-changing tidal potential and mass-transfer rate and the
inclination angle on the structure and evolution of the accretion
flow. Another important characteristic of Be/X-ray binaries is that
they are systems with double circumstellar discs (the Be disc and the
accretion disc). The interaction in such systems is threefold: the
interactions between the neutron star and the Be disc, the Be star and
the accretion disc and the Be disc and the accretion disc (mainly via
the mass transfer). Much more work is desirable both theoretically and
observationally in order to understand this interesting and important 
group of objects.

\section*{Acknowledgements}

We acknowlege the anonymous refree for constructive comments.
We would like to thank Matthew R.\ Bate for allowing us to use his SPH code. 
KH acknowledes James R. Murray for useful comments during the short
stay at the Swinburne University of Technology, Australia. 
KH also thanks Noboru Kaneko for his continuous encouragement.
The simulations reported here were performed using the facility at the
Hokkaido University Information Initiative Center, Japan. This work
was supported in part by Grant-in-Aid for Scientific Reserch
(13640244) of Japan Society for the Promotion of Science.

\appendix

\end{document}